\documentclass[aps,prl,reprint,superscriptaddress]{revtex4-2}
\usepackage{amsmath}
\usepackage{amsfonts}
\usepackage{amssymb}
\usepackage{braket}
\usepackage{tcolorbox}
\usepackage{xcolor}
\usepackage{graphicx}
\usepackage{enumerate}
\definecolor{boxback}{RGB}{246, 246, 246}

\begin{document}

\title{Reaching states below the threshold energy in spin glasses via quantum annealing}

\author{Christopher~L.~Baldwin}
\email{baldw292@msu.edu}
\affiliation{Department of Physics and Astronomy, Michigan State University, East Lansing, Michigan 48824, USA}

\begin{abstract}
Although quantum annealing is usually considered as a method for locating the ground states of difficult spin-glass and optimization problems, its use in \textit{approximate} optimization --- finding low- but not zero-energy states in a reasonably short amount of time --- is no less important.
Here we investigate the behavior of quantum annealing at approximate optimization in the canonical mean-field spin-glass models, the spherical $p$-spin models, and find that it performs surprisingly well.
Whereas it had long been assumed that infinite-range spin glasses have a unique ``threshold'' energy at which all quench and annealing dynamics become trapped until exponential timescales, recent work has shown that two-stage quenches can in fact reach states below the naive threshold in more generic situations.
We demonstrate that quantum annealing is also capable of exploiting this effect to locate sub-threshold states in $O(1)$ time.
Not only can it attain energies as far below the threshold as classical annealing algorithms, but it can do so significantly faster: for an annealing schedule taking time $\tau$, the residual energy under quantum annealing decays as $\tau^{-\alpha}$ with an exponent up to twice as large as that of simulated annealing in the cases considered.
Importantly, by deriving and numerically solving closed integro-differential equations that hold in the thermodynamic limit, our results are free from finite-size effects and hold for annealing times that are unambiguously independent of system size.
\end{abstract}

\maketitle


\textit{Introduction}---Quantum annealing (QA) is usually studied as a technique for finding the ground states of difficult spin-glass and optimization problems~\cite{Albash2018Adiabatic,Hauke2020Perspectives,Rajak2023Quantum}.
Since many problems of practical significance can be represented in this framework, ranging from logistics~\cite{Neukart2017Traffic,Weinberg2023Supply} to finance~\cite{Orus2019Forecasting,Mugel2021Hybrid} to medicine~\cite{Li2018QuantumII,Boev2021Genome,Irback2022Folding,Gircha2023Hybrid}, understanding whether and how QA can provide a quantum advantage has been an important and long-standing research goal.

Typically, one seeks the ground state of a classical Ising Hamiltonian $H_0$ (involving $N$ spins) and applies a transverse field with strength controlled by a parameter $s$:
\begin{equation} \label{eq:quantum_annealing_Hamiltonian}
H(s) = s H_0 \big( \hat{\sigma}^z \big) - (1 - s) \sum_{j=1}^N \hat{\sigma}_j^x.
\end{equation}
The spins are initialized in the product state aligned with the field, and left to evolve under $H(s)$ while increasing $s$ from 0 to 1 --- the adiabatic theorem implies that the spins will end up in the ground state of $H_0$ if $s$ is varied sufficiently slowly~\cite{Messiah1962,Jansen2007Bounds}.
Unfortunately, a number of works have established that the timescale required to remain in the ground state is generically exponential in $N$, since the same ``rugged energy landscape'' that impedes classical algorithms gives rise to numerous exponentially small gaps within the spin-glass phase~\cite{Jorg2008Simple,Altshuler2010Anderson,Foini2010Solvable,Bapst2013Quantum,Knysh2016ZeroTemperature,Baldwin2018Quantum}.
Thus there are significant obstacles to using QA as a means to efficiently locate ground states, especially in the current NISQ era, where experimental quantum annealers are limited to short annealing times~\cite{King2022Coherent,King2023Quantum,King2025BeyondClassical}.

That said, QA could still be useful for \textit{approximate} optimization: carry out the annealing over a reasonably short time and see if it yields lower-energy states than could be obtained by other methods in comparable time.
This has been explored far less than the ground-state question, yet some recent works on the topic have been quite promising~\cite{Braida2024Tight,Zhang2024Cyclic,MunozBauza2025Scaling}.
Thus there is a need for a more systematic investigation, particularly via controlled analytical studies that do not suffer from finite-size effects while still applying to genuinely hard problems.

Here we carry out this investigation using the canonical $p$-spin models of spin glasses~\cite{Crisanti1992Spherical,Crisanti1993Spherical,Castellani2005Spin}, and find that in certain regimes, QA is superior to simulated annealing (SA) at approximate optimization.
While SA is not usually considered state-of-the-art, it is a natural first comparison when assessing the performance of QA.
Our results thus provide a valuable theoretical basis for applying QA to approximate optimization problems.

The $p$-spin models have long played an important role in the spin-glass field~\cite{Mezard1987,Fischer1991,Nishimori2001,Mezard2009}, particularly at the intersection with theoretical computer science and information theory, since they exhibit the phenomenology expected from more realistic problems --- exponentially many local minima, extensive energy barriers, exponential time to reach low-energy states --- while remaining solvable by analytic means.
In particular, one can derive closed equations for the correlation and response functions, as well as the average energy, under various types of dynamics in the thermodynamic limit~\cite{Cugliandolo1993Analytical,Barrat1996Dynamics,Cugliandolo1999RealTime,Cugliandolo2017NonEquilibrium}.
This allows us to avoid the finite-size effects that plague numerical studies of QA while still analyzing a highly frustrated many-body model.

However, until recently, the theory of spin glasses would have discouraged using QA for approximate optimization.
A central concept is that of the ``threshold'' energy~\cite{Crisanti1995ThoulessAndersonPalmer,Castellani2005Spin,Bray2007Statistics}, which is (loosely speaking) the energy at which almost all local minima of the energy landscape lie.
It stands to reason that any SA protocol, regardless of schedule~\cite{terminology_note}, would become trapped in one of those minima --- thus SA is capable of reaching the threshold energy but no lower on any sub-exponential timescale.
Presumably the wavefunction under QA would be trapped by the same minima, meaning QA can also reach the threshold energy but no lower, and there would be no advantage to using QA over SA.

Refs.~\cite{Folena2020Rethinking,Folena2021Gradient} have recently disproven the notion of a unique threshold energy in ``mixed'' $p$-spin models (which, despite the relatively less attention paid to them, are arguably more generic).
The authors showed that while a naive quench (i.e., steepest descent in algorithmic terms) does indeed approach the putative threshold energy, a two-stage quench --- in which the system is first thermalized at an intermediate temperature --- is capable of reaching unambiguously lower energies in $O(1)$ time.

Since different SA protocols can reach different energies in the mixed models, it is natural to ask what energies QA can achieve.
We answer this question here.
We find that QA is capable of reaching energies as far below the threshold as SA, which is already surprising given the variation among SA protocols.
More importantly, we observe that QA can in some cases reach this optimal sub-threshold energy more rapidly than SA: for a protocol taking $O(1)$ time $\tau$, the residual energy above the asymptotic value decays as a power law $\tau^{-\alpha}$ under both SA and QA, but with an exponent $\alpha$ that can be up to twice as large (and likely more) for QA.
In such situations, QA has a clear advantage over SA in reaching sub-threshold energies.
In what follows, we elaborate on and clarify these claims.


\textit{Models}---The mixed $p$-spin model is given by the Hamiltonian
\begin{equation} \label{eq:p_spin_Hamiltonian}
H_0(\sigma) = \sum_{p=2}^{\infty} \sqrt{a_p} \sum_{j_1 < \cdots < j_p} J_{j_1 \cdots j_p} \sigma_{j_1} \cdots \sigma_{j_p}.
\end{equation}
In order to study various types of dynamics analytically, we must consider a spherical version of the model (as is common in the field): each $\sigma_j$ is a continuous variable subject only to the spherical constraint $\sum_j \sigma_j^2 = N$.
Thus we equivalently interpret Eq.~\eqref{eq:p_spin_Hamiltonian} as a potential energy landscape on the surface of an $N$-dimensional hypersphere of radius $\sqrt{N}$.
Each coefficient $J_{j_1 \cdots j_p}$ is an independent Gaussian random variable of mean zero and variance $p!/2N^{p-1}$.
We calculate the average values of quantities with respect to the coefficients (using $\mathbb{E}[ \, \cdot \, ]$ to denote the average).
The coefficients $a_p$, meanwhile, are fixed parameters specifying the model.

A direct calculation shows that the covariance between two points $\sigma \equiv \{ \sigma_j \}$ and $\sigma' \equiv \{ \sigma'_j \}$ on the sphere is, up to terms which are subleading as $N \rightarrow \infty$,
\begin{equation} \label{eq:p_spin_covariance}
\mathbb{E} \big[ H_0(\sigma) H_0(\sigma') \big] \sim \frac{N}{2} f \Big[ N^{-1} \sum_j \sigma_j \sigma_j' \Big],
\end{equation}
where $f[Q] \equiv \sum_p a_p Q^p$.
It is usually simpler to specify the model by giving the polynomial $f[Q]$ directly.
The ``pure'' $p$-spin model corresponds to $f[Q] = Q^p$.
The results of Ref.~\cite{Folena2020Rethinking} are for $f[Q] = Q^3 + Q^4$, and here we generalize to $f[Q] = Q^3 + Q^p$.


We assess the performance of both SA and QA in approximate optimization of the $p$-spin model by calculating the average energy density $\epsilon(\tau) \equiv \mathbb{E}[\langle H_0(\tau) \rangle]/N$ at the end of a protocol taking time $\tau$.
To make the notion of ``reasonably short'' annealing times precise, we focus on how $\epsilon(\tau)$ behaves at $\tau$ which are large but $O(1)$ with respect to $N$.

For SA, we study the Langevin dynamics of the model:
\begin{equation} \label{eq:Langevin_dynamics}
\partial_t \sigma_j(t) = -s(t) \frac{\partial H_0}{\partial \sigma_j} + \sqrt{1 - s(t)} \xi_j(t) - z(t) \sigma_j(t),
\end{equation}
where $\xi_j(t)$ is white noise with mean zero and covariance $\langle \xi_j(t) \xi_{j'}(t') \rangle = 2 \delta_{jj'} \delta(t - t')$ (here $\langle \, \cdot \, \rangle$ denotes an average over noise, distinct from the average $\mathbb{E}[ \, \cdot \, ]$ over coefficients in $H_0$).
The parameter $s$, analogous to Eq.~\eqref{eq:quantum_annealing_Hamiltonian}, controls the relative strength of the energy landscape and fluctuations (note that the effective temperature of the noise is $(1 - s)/s$).
We increase $s$ from 0 to 1 (infinite to zero temperature) over the course of the annealing protocol, although we use and compare different time-dependences for $s(t)$.
Lastly, the final term in Eq.~\eqref{eq:Langevin_dynamics} is to enforce the spherical constraint: the time-dependence of $z(t)$ is chosen to ensure that $\sum_j \langle \sigma_j(t)^2 \rangle = N$ at all times.

Using a path-integral representation of Eq.~\eqref{eq:Langevin_dynamics}, equations for the correlation function $C(t, t')$ and response function $R(t, t')$ can be derived.
This is well-documented in the literature (see Refs.~\cite{Castellani2005Spin,Folena2020Rethinking} and references therein), so we give only the final result:
\begin{widetext}
\begin{equation} \label{eq:Langevin_correlation_equation}
\partial_t C(t, t') = -z(t) C(t, t') + \frac{s(t)}{2} \int_0^t \textrm{d}t'' s(t'') f'' \big[ C(t, t'') \big] C(t'', t') R(t, t'') + \frac{s(t)}{2} \int_0^{t'} \textrm{d}t'' s(t'') f' \big[ C(t, t'') \big] R(t', t''),
\end{equation}
\begin{equation} \label{eq:Langevin_response_equation}
\partial_t R(t, t') = -z(t) R(t, t') + \frac{s(t)}{2} \int_{t'}^t \textrm{d}t'' s(t'') f'' \big[ C(t, t'') \big] R(t, t'') R(t'', t'),
\end{equation}
with $z(t)$ given by
\begin{equation} \label{eq:Langevin_Lagrange_equation}
z(t) = \frac{s(t)}{2} \int_0^t \textrm{d}t'' s(t'') f'' \big[ C(t, t'') \big] C(t, t'') R(t, t'') + \frac{s(t)}{2} \int_0^t \textrm{d}t'' s(t'') f' \big[ C(t, t'') \big] R(t, t'') + 1 - s(t).
\end{equation}
\end{widetext}
The average energy density at time $t$ is
\begin{equation} \label{eq:Langevin_energy_equation}
\epsilon(t) = -\frac{1}{2} \int_0^t \textrm{d}t'' s(t'') f' \big[ C(t, t'') \big] R(t, t'').
\end{equation}
These equations hold in the $N \rightarrow \infty$ limit, for all times that are $O(1)$ with respect to $N$.
We assume $t \geq t'$, since $C(t', t) = C(t, t')$ and $R(t', t) = 0$ by causality.
In all equations, $f[Q]$ is the same polynomial as defined below Eq.~\eqref{eq:p_spin_covariance}, and primes denote its derivatives.

Eqs.~\eqref{eq:Langevin_correlation_equation} through~\eqref{eq:Langevin_Lagrange_equation} must be solved numerically, which requires introducing a temporal discretization $\Delta t$.
The discretization is straightforward: replace $\partial_t$ by the standard forward finite-difference and $\int \textrm{d}t''$ by $\sum_{t''} \Delta t$.
Initial conditions are that $C(t, t) = 1$ and $R(t + \Delta t, t) = 1$.
Note that finite $\Delta t$ is the only source of error in our calculations, and we have confirmed that our results (in particular the exponents $\alpha$ below) are insensitive to the values of $\Delta t$ that we use.

For QA, we replace each classical variable $\sigma_j$ with a position operator $\hat{X}_j$ and introduce a conjugate momentum $\hat{P}_j$.
Analogous to Eq.~\eqref{eq:quantum_annealing_Hamiltonian}, the system evolves under Hamiltonian
\begin{equation} \label{eq:Hamiltonian_dynamics}
H(t) = s(t) H_0 \big( \hat{X} \big) + \frac{1 - s(t)}{2} \sum_j \hat{P}_j^2 + \frac{z(t)}{2} \sum_j \hat{X}_j^2,
\end{equation}
with the kinetic energy playing the role of a transverse field~\cite{transverse_field_note} and $s(t)$ increasing from 0 to 1.
Once again, $z(t)$ is chosen to ensure that $\sum_j \langle \hat{X}_j(t)^2 \rangle = N$ at all times (here $\langle \, \cdot \, \rangle$ denotes the quantum-mechanical expectation value, still distinct from $\mathbb{E}[ \, \cdot \, ]$).

Equations for the correlation function $C(t, t')$ and response function $R(t, t')$ can again be derived by using the appropriate path integral:
\begin{widetext}
\begin{equation} \label{eq:Hamiltonian_correlation_equation}
\partial_t \Big( \big( 1 - s(t) \big)^{-1} \partial_t C(t, t') \Big) = -z(t) C(t, t') - s(t) \int_0^t \textrm{d}t'' s(t'') \textrm{Im} f' \Big[ C(t, t'') - \frac{i}{2} R(t, t'') \Big] \Big( C(t', t'') - \frac{i}{2} R(t', t'') \Big),
\end{equation}
\begin{equation} \label{eq:Hamiltonian_response_equation}
\partial_t \Big( \big( 1 - s(t) \big)^{-1} \partial_t R(t, t') \Big) = -z(t) R(t, t') - s(t) \int_{t'}^t \textrm{d}t'' s(t'') \textrm{Im} f' \Big[ C(t, t'') - \frac{i}{2} R(t, t'') \Big] R(t'', t'),
\end{equation}
with $z(t)$ given by
\begin{equation} \label{eq:Hamiltonian_Lagrange_equation}
z(t) = -\big( 1 - s(t) \big)^{-1} \partial_t^2 C(t, t') \Big|_{t'=t} - s(t) \int_0^t \textrm{d}t'' s(t'') \textrm{Im} f' \Big[ C(t, t'') - \frac{i}{2} R(t, t'') \Big] \Big( C(t, t'') - \frac{i}{2} R(t, t'') \Big).
\end{equation}
\end{widetext}
The average energy density at time $t$ is
\begin{equation} \label{eq:Hamiltonian_energy_equation}
\epsilon(t) = \int_0^t \textrm{d}t'' s(t'') \textrm{Im} f \Big[ C(t, t'') - \frac{i}{2} R(t, t'') \Big].
\end{equation}
These equations as well have been used a number of times in the literature~\cite{Cugliandolo1999RealTime,Cugliandolo2019Role,Thomson2020Quantum}, but since the appropriate discretization is more subtle, we give a self-contained derivation with the correct discretization in the Supplement.


\textit{Threshold energy}---Before presenting results, we briefly discuss the concept of the threshold energy more precisely.
Still interpreting $H_0(\sigma)$ as a potential energy surface on the hypersphere, one focuses on the stationary points, i.e., points where $\nabla H_0(\sigma) = 0$.
The eigenvalues of the Hessian at a stationary point indicate whether it is a local minimum or a saddle point: all eigenvalues are positive in the former and some are negative in the latter.
In the \textit{pure} $p$-spin models, the eigenvalue distribution at a stationary point is a shifted semicircle whose center $\mu$ is uniquely determined by the energy density $\epsilon$.
There is a critical value $\epsilon_{\textrm{th}} = -\sqrt{2(p-1)/p}$ such that stationary points are minima for $\epsilon < \epsilon_{\textrm{th}}$ and saddles for $\epsilon > \epsilon_{\textrm{th}}$.
Furthermore, the number of stationary points at a given energy can be computed --- it is found to scale as $e^{N \Sigma(\epsilon)}$, with exponent $\Sigma(\epsilon)$ that increases monotonically for $\epsilon < \epsilon_{\textrm{th}}$.
In other words, all but an exponentially small fraction of local minima are located at the threshold energy.
As discussed above, it would be reasonable to assume that this is the energy at which all quench and annealing dynamics become trapped.

In the \textit{mixed} $p$-spin models, the eigenvalue distribution is still a semicircle, but the center $\mu$ is no longer fixed by the energy $\epsilon$ --- stationary points with macroscopically distinct eigenvalue distributions coexist at the same energy.
Nonetheless, one can compute the number of stationary points as a function of $\mu$ and $\epsilon$ together, finding that it still scales exponentially as $e^{N \Sigma(\epsilon, \mu)}$.
Thus despite the coexistence of different stationary points, all but an exponentially small fraction at a given energy do have a specific value of $\mu$.
One then defines the threshold energy as where \textit{typical} stationary points transition from being minima to saddles, and it is known to have the value~\cite{Auffinger2013Complexity}
\begin{equation} \label{eq:mixed_threshold_energy}
\epsilon_{\textrm{th}} = -\frac{f[1] \big( f''[1] - f'[1] \big) + f'[1]^2}{f'[1] \sqrt{2 f''[1]}}.
\end{equation}


\textit{Results}---We compare the energy obtained at the end of an SA protocol (Eqs.~\eqref{eq:Langevin_correlation_equation} through~\eqref{eq:Langevin_energy_equation}) with the energy obtained at the end of a QA protocol (Eqs.~\eqref{eq:Hamiltonian_correlation_equation} through~\eqref{eq:Hamiltonian_energy_equation}).
Denote the protocol runtime by $\tau$ and the final energy by $\epsilon(\tau)$.
To reiterate, by determining $\epsilon(\tau)$ through the integro-differential equations above, we are automatically considering $\tau$ that are independent of system size.
We study multiple SA protocols: i) a naive quench, with $s(t) = 1$ for all $t \leq \tau$; ii) a two-stage quench, with $s(t) = s_0$ for $t < \tau/2$ and $s(t) = 1$ for $t > \tau/2$; iii) an anneal, with $s(t) = t/\tau$.
For the QA protocol, we only study the anneal, $s(t) = t/\tau$.

\begin{figure}[t]
\centering
\includegraphics[width=1.0\columnwidth]{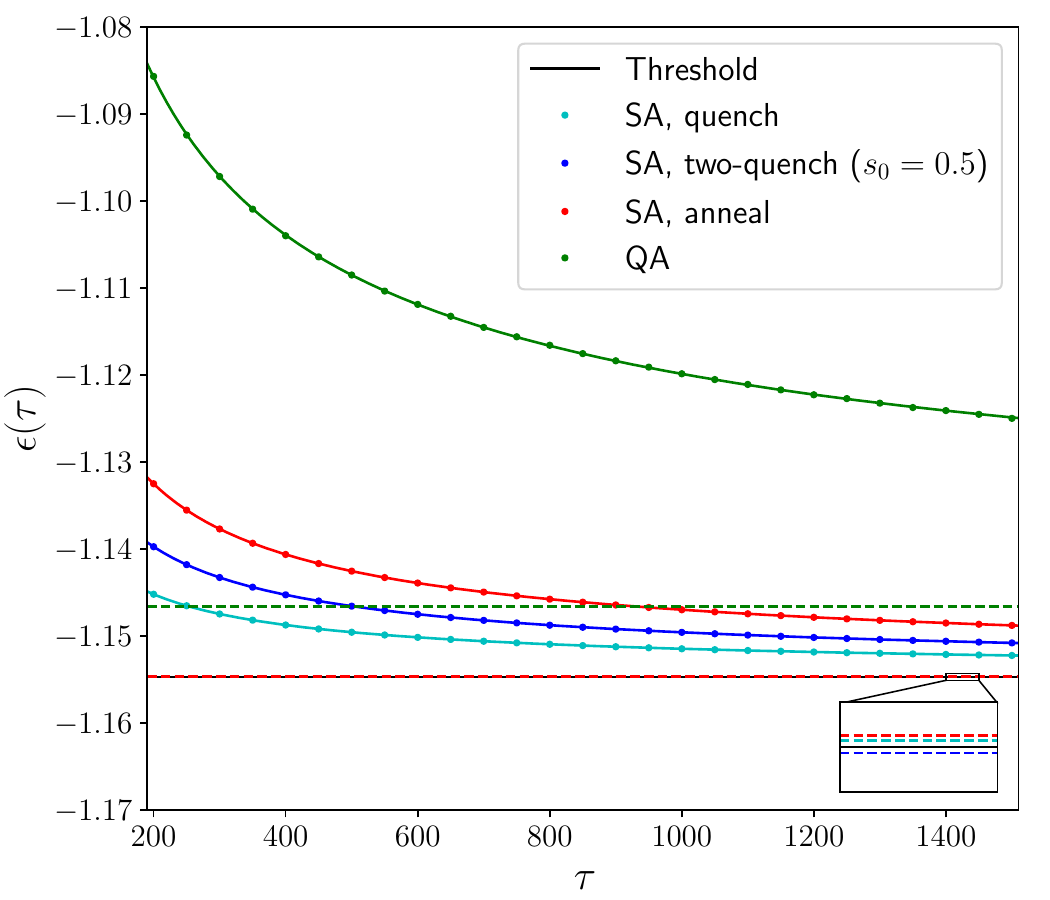}
\caption{Average energy density $\epsilon(\tau)$ at the end of various protocols as a function of protocol runtime $\tau$, for the pure model $f[Q] = Q^3$. Each set of points is fit to a power-law decay (Eq.~\eqref{eq:power_law_decay}), with the fitted curve shown as the solid colored line. The large-$\tau$ limit of the fitted curve is indicated by the dashed colored line, and the solid black line indicates the threshold energy $\epsilon_{\textrm{th}}$ (Eq.~\eqref{eq:mixed_threshold_energy}). Calculations for all protocols use $\Delta t = 0.1$.}
\label{fig:pure_p_spin}
\end{figure}

As a point of comparison, consider the performance of each protocol in the pure $p$-spin model, e.g., $p = 3$ in Fig.~\ref{fig:pure_p_spin} (the conclusions are the same for other $p$).
For each protocol, we fit $\epsilon(\tau)$ to a power-law decay:
\begin{equation} \label{eq:power_law_decay}
\epsilon(\tau) \approx \epsilon_{\infty} + C \tau^{-\alpha}.
\end{equation}
The solid colored lines in Fig.~\ref{fig:pure_p_spin} show the fitted curves, which match the data quite well in all cases, and the dashed colored lines show the asymptotic values $\epsilon_{\infty}$.
The asymptotic values for the SA protocols agree remarkably well (to the fourth decimal place) with the threshold energy $\epsilon_{\textrm{th}} = -2/\sqrt{3}$ shown in black (this can be proven analytically for the naive quench but not for the others~\cite{Cugliandolo1993Analytical}).
Although the asymptotic value for QA disagrees with $\epsilon_{\textrm{th}}$ by slightly less than $1\%$, this is likely an artifact of the fit~\cite{fitting_note}.
As for the exponents $\alpha$, we find that $\alpha \approx 0.51$ for QA and $\alpha \approx 0.66$ for all SA protocols.
Thus overall, QA performs quite poorly in the pure $p$-spin model --- not only does it fail to reach lower energies than the SA protocols, becoming trapped at the same $\epsilon_{\textrm{th}}$, but it is slower to reach that value.

\begin{figure}[t]
\centering
\includegraphics[width=1.0\columnwidth]{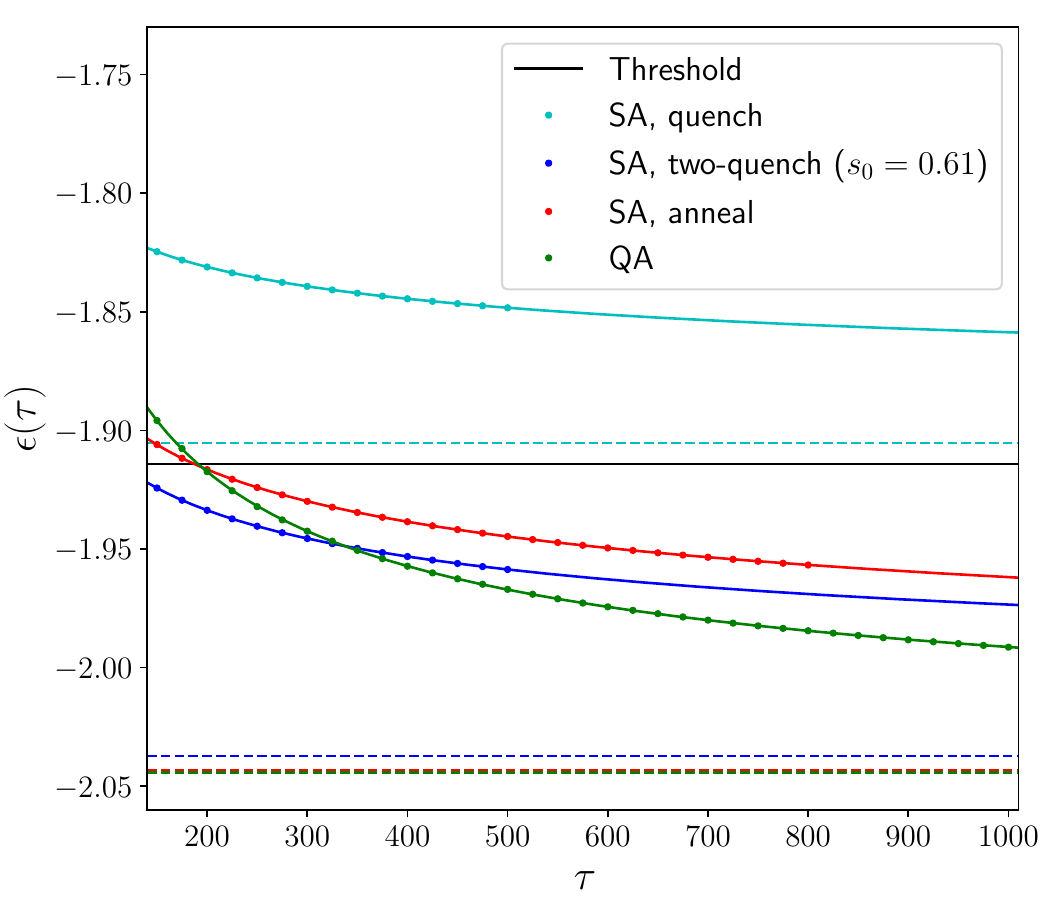}
\caption{Average energy density $\epsilon(\tau)$ at the end of various protocols as a function of protocol runtime $\tau$, for the mixed model $f[Q] = Q^3 + Q^{14}$. Each set of points is fit to a power-law decay (Eq.~\eqref{eq:power_law_decay}), with the fitted curve shown as the solid colored line. The large-$\tau$ limit of the fitted curve is indicated by the dashed colored line, and the solid black line indicates the threshold energy $\epsilon_{\textrm{th}}$ (Eq.~\eqref{eq:mixed_threshold_energy}). Calculations for the quench and two-stage quench use $\Delta t = 0.02$, while those for the anneals use $\Delta t = 0.04$.}
\label{fig:mixed_p_spin}
\end{figure}

Contrast with the performance in mixed models, e.g., $f[Q] = Q^3 + Q^{14}$ in Fig.~\ref{fig:mixed_p_spin}, where the differences are especially pronounced.
Consistent with Refs.~\cite{Folena2020Rethinking,Folena2021Gradient}, the naive quench decays to a value quite close to the threshold energy (black line).
Two-stage quenches are capable of reaching unambiguously lower energies, however~\cite{Folena2020Rethinking,Folena2021Gradient} --- the value of $s_0$ used in Fig.~\ref{fig:mixed_p_spin} is that which gives the lowest asymptotic energy, for which $\epsilon(\tau)$ is already well below $\epsilon_{\textrm{th}}$ even at accessible $\tau$.
We find that the quantum and classical anneals both reach below the threshold as well, with asymptotic values that are very close to that of the optimal two-stage quench.
Given the range of energies that different SA protocols can reach, this is non-trivial in of itself.

\begin{figure}[t]
\centering
\includegraphics[width=1.0\columnwidth]{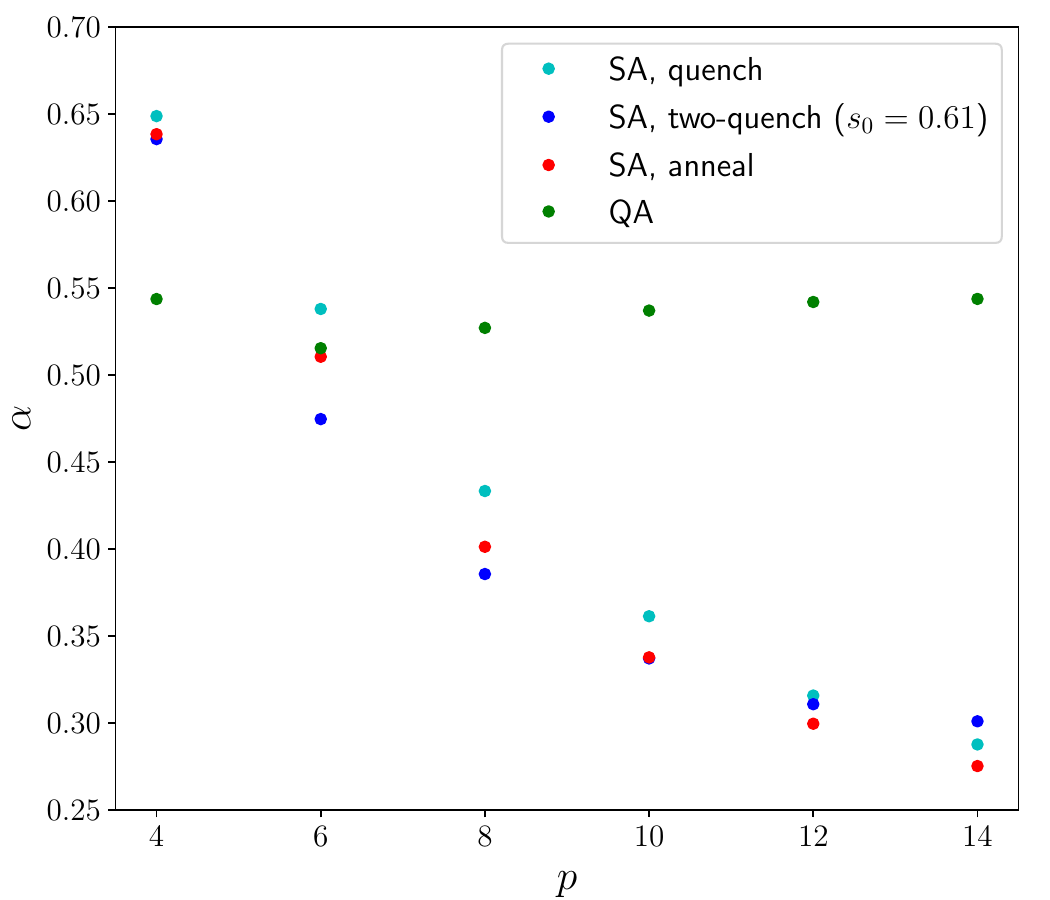}
\caption{Fitted exponents $\alpha$ (Eq.~\eqref{eq:power_law_decay}) for various protocols in the mixed model $f[Q] = Q^3 + Q^p$, as a function of $p$. Calculations for the quench and two-stage quench use timestep $\Delta t = 0.02$, while those for the anneals use $\Delta t = 0.04$.}
\label{fig:exponents}
\end{figure}

Moreover, Fig.~\ref{fig:mixed_p_spin} makes clear that QA approaches the asymptotic energy \textit{faster} than the SA protocols.
The power-law decay for QA has exponent $\alpha \approx 0.54$, while the classical anneal has $\alpha \approx 0.28$ and the two-stage quench has $\alpha \approx 0.30$ (see Fig.~\ref{fig:exponents}).
Thus QA performs objectively \textit{better} than SA in this specific mixed model: it reaches energies as low as the classical protocols can and does so with a significantly faster power-law decay.

Fig.~\ref{fig:exponents} shows how the exponents $\alpha$ vary with $p$ in the mixed model $f[Q] = Q^3 + Q^p$.
For $p = 4$, like in the pure model, QA has a smaller exponent than the SA protocols (thus QA does not always outperform SA even in mixed models).
Yet whereas $\alpha$ falls off rapidly with $p$ in SA, it remains roughly independent of $p$ in QA.
By $p = 14$, QA has an exponent roughly twice as large as the SA protocols, and the ratio would likely increase further were we to increase $p$~\cite{p_limit_note}.

This behavior offers a clue as to why, and in which models, QA has an advantage over SA.
As $p$ increases, the $p$-body terms in the energy landscape become highly ``spiked'', in that the energies at different points become uncorrelated and wells in the landscape become narrow.
In mixed models, the spikes caused by high-$p$ terms may act as obstacles that impede descent towards the broader minima produced by low-$p$ terms.
Our results would then suggest that QA is better at avoiding these traps.
This is reasonable to suspect, not only because of the quantum tunneling that is usually discussed as an advantage of QA (at least for escaping from sufficiently shallow local minima), but also because the Hamiltonian dynamics of QA does not explicitly follow the local gradient of the landscape, whereas SA does.
Yet this is purely speculative --- further investigation is certainly warranted.


\textit{Conclusion}---We have compared the performance of QA to that of SA in approximate optimization of the spherical $p$-spin models, which have long been prominent in the development of spin-glass theory.
We have found that in certain cases, QA in fact outperforms SA: it reaches energies as low as the latter does and with a significantly faster power-law decay (specifically for times that are $O(1)$ with respect to system size).
Although QA does not always have an advantage over SA, the cases in which it does are in no way fine-tuned.
Unlike the vast majority of QA studies, these results are immune to finite-size effects and do not require any extrapolation from small-size numerical data.
The results both provide a theoretical foundation for using QA in approximate optimization and identify promising directions in which to look for practical quantum advantages.

That said, there are a number of important questions that remain to be investigated.
Admittedly, the basic SA that we have considered here is not a state-of-the-art classical algorithm --- many specific problems have dedicated solvers, and there are more sophisticated general-purpose algorithms such as belief propagation and parallel tempering.
QA should be compared to these methods as well, although analytical studies analogous to what we have done here would likely be quite difficult.

Since present-day quantum annealers are designed for discrete-variable optimization, the behavior of QA in the Ising versions of the $p$-spin models should also be considered (although this as well would be difficult to do analytically).
It is reasonable to expect that QA would have a similar advantage over SA in the Ising case, given the similarity in the qualitative physics of the Ising and spherical models, but this is not guaranteed.

Lastly, the physics underlying our results would benefit from further study, particularly the role of quantum mechanics, as this would inform whether alternate classical algorithms might have a similar advantage.
We have speculated that the superiority of QA is due to it better avoiding narrow traps in the energy landscape.
If this derives merely from the Hamiltonian nature of QA dynamics (which is unrelated to quantum mechanics), then classical Hamiltonian-based methods may perform just as well~\cite{Neal2011MCMC,Betancourt2017Geometric} (the latter are well-studied for continuous-variable optimization problems, although whether they can be generalized to discrete-variable problems without costs to runtime scaling is non-trivial~\cite{vector_spin_note}).
Yet if quantum tunneling plays a necessary role, then QA would have a genuine advantage over classical algorithms.


\textit{Acknowledgements}---It is a pleasure to thank Anushya Chandran, Philip Crowley, and Sumner Hearth for valuable discussions and feedback.
This work was supported by the U.S. National Science Foundation under award No.~2508604.

\bibliography{biblio}

\begin{thebibliography}{50}%
\makeatletter
\providecommand \@ifxundefined [1]{%
 \@ifx{#1\undefined}
}%
\providecommand \@ifnum [1]{%
 \ifnum #1\expandafter \@firstoftwo
 \else \expandafter \@secondoftwo
 \fi
}%
\providecommand \@ifx [1]{%
 \ifx #1\expandafter \@firstoftwo
 \else \expandafter \@secondoftwo
 \fi
}%
\providecommand \natexlab [1]{#1}%
\providecommand \enquote  [1]{``#1''}%
\providecommand \bibnamefont  [1]{#1}%
\providecommand \bibfnamefont [1]{#1}%
\providecommand \citenamefont [1]{#1}%
\providecommand \href@noop [0]{\@secondoftwo}%
\providecommand \href [0]{\begingroup \@sanitize@url \@href}%
\providecommand \@href[1]{\@@startlink{#1}\@@href}%
\providecommand \@@href[1]{\endgroup#1\@@endlink}%
\providecommand \@sanitize@url [0]{\catcode `\\12\catcode `\$12\catcode `\&12\catcode `\#12\catcode `\^12\catcode `\_12\catcode `\%12\relax}%
\providecommand \@@startlink[1]{}%
\providecommand \@@endlink[0]{}%
\providecommand \url  [0]{\begingroup\@sanitize@url \@url }%
\providecommand \@url [1]{\endgroup\@href {#1}{\urlprefix }}%
\providecommand \urlprefix  [0]{URL }%
\providecommand \Eprint [0]{\href }%
\providecommand \doibase [0]{https://doi.org/}%
\providecommand \selectlanguage [0]{\@gobble}%
\providecommand \bibinfo  [0]{\@secondoftwo}%
\providecommand \bibfield  [0]{\@secondoftwo}%
\providecommand \translation [1]{[#1]}%
\providecommand \BibitemOpen [0]{}%
\providecommand \bibitemStop [0]{}%
\providecommand \bibitemNoStop [0]{.\EOS\space}%
\providecommand \EOS [0]{\spacefactor3000\relax}%
\providecommand \BibitemShut  [1]{\csname bibitem#1\endcsname}%
\let\auto@bib@innerbib\@empty
\bibitem [{\citenamefont {Albash}\ and\ \citenamefont {Lidar}(2018)}]{Albash2018Adiabatic}%
  \BibitemOpen
  \bibfield  {author} {\bibinfo {author} {\bibfnamefont {T.}~\bibnamefont {Albash}}\ and\ \bibinfo {author} {\bibfnamefont {D.~A.}\ \bibnamefont {Lidar}},\ }\bibfield  {title} {\bibinfo {title} {Adiabatic quantum computation},\ }\href@noop {} {\bibfield  {journal} {\bibinfo  {journal} {Rev. Mod. Phys.}\ }\textbf {\bibinfo {volume} {90}},\ \bibinfo {pages} {015002} (\bibinfo {year} {2018})}\BibitemShut {NoStop}%
\bibitem [{\citenamefont {Hauke}\ \emph {et~al.}(2020)\citenamefont {Hauke}, \citenamefont {Katzgraber}, \citenamefont {Lechner}, \citenamefont {Nishimori},\ and\ \citenamefont {Oliver}}]{Hauke2020Perspectives}%
  \BibitemOpen
  \bibfield  {author} {\bibinfo {author} {\bibfnamefont {P.}~\bibnamefont {Hauke}}, \bibinfo {author} {\bibfnamefont {H.~G.}\ \bibnamefont {Katzgraber}}, \bibinfo {author} {\bibfnamefont {W.}~\bibnamefont {Lechner}}, \bibinfo {author} {\bibfnamefont {H.}~\bibnamefont {Nishimori}},\ and\ \bibinfo {author} {\bibfnamefont {W.~D.}\ \bibnamefont {Oliver}},\ }\bibfield  {title} {\bibinfo {title} {Perspectives of quantum annealing: {M}ethods and implementations},\ }\href@noop {} {\bibfield  {journal} {\bibinfo  {journal} {Rep. Prog. Phys.}\ }\textbf {\bibinfo {volume} {83}},\ \bibinfo {pages} {054401} (\bibinfo {year} {2020})}\BibitemShut {NoStop}%
\bibitem [{\citenamefont {Rajak}\ \emph {et~al.}(2023)\citenamefont {Rajak}, \citenamefont {Suzuki}, \citenamefont {Dutta},\ and\ \citenamefont {Chakrabarti}}]{Rajak2023Quantum}%
  \BibitemOpen
  \bibfield  {author} {\bibinfo {author} {\bibfnamefont {A.}~\bibnamefont {Rajak}}, \bibinfo {author} {\bibfnamefont {S.}~\bibnamefont {Suzuki}}, \bibinfo {author} {\bibfnamefont {A.}~\bibnamefont {Dutta}},\ and\ \bibinfo {author} {\bibfnamefont {B.~K.}\ \bibnamefont {Chakrabarti}},\ }\bibfield  {title} {\bibinfo {title} {Quantum annealing: {A}n overview},\ }\href@noop {} {\bibfield  {journal} {\bibinfo  {journal} {Phil. Trans. R. Soc. A}\ }\textbf {\bibinfo {volume} {381}},\ \bibinfo {pages} {20210417} (\bibinfo {year} {2023})}\BibitemShut {NoStop}%
\bibitem [{\citenamefont {Neukart}\ \emph {et~al.}(2017)\citenamefont {Neukart}, \citenamefont {Compostella}, \citenamefont {Seidel}, \citenamefont {von Dollen}, \citenamefont {Yarkoni},\ and\ \citenamefont {Parney}}]{Neukart2017Traffic}%
  \BibitemOpen
  \bibfield  {author} {\bibinfo {author} {\bibfnamefont {F.}~\bibnamefont {Neukart}}, \bibinfo {author} {\bibfnamefont {G.}~\bibnamefont {Compostella}}, \bibinfo {author} {\bibfnamefont {C.}~\bibnamefont {Seidel}}, \bibinfo {author} {\bibfnamefont {D.}~\bibnamefont {von Dollen}}, \bibinfo {author} {\bibfnamefont {S.}~\bibnamefont {Yarkoni}},\ and\ \bibinfo {author} {\bibfnamefont {B.}~\bibnamefont {Parney}},\ }\bibfield  {title} {\bibinfo {title} {Traffic flow optimization using a quantum annealer},\ }\href@noop {} {\bibfield  {journal} {\bibinfo  {journal} {Front. ICT}\ }\textbf {\bibinfo {volume} {4}},\ \bibinfo {pages} {29} (\bibinfo {year} {2017})}\BibitemShut {NoStop}%
\bibitem [{\citenamefont {Weinberg}\ \emph {et~al.}(2023)\citenamefont {Weinberg}, \citenamefont {Sanches}, \citenamefont {Ide}, \citenamefont {Kamiya},\ and\ \citenamefont {Correll}}]{Weinberg2023Supply}%
  \BibitemOpen
  \bibfield  {author} {\bibinfo {author} {\bibfnamefont {S.~J.}\ \bibnamefont {Weinberg}}, \bibinfo {author} {\bibfnamefont {F.}~\bibnamefont {Sanches}}, \bibinfo {author} {\bibfnamefont {T.}~\bibnamefont {Ide}}, \bibinfo {author} {\bibfnamefont {K.}~\bibnamefont {Kamiya}},\ and\ \bibinfo {author} {\bibfnamefont {R.}~\bibnamefont {Correll}},\ }\bibfield  {title} {\bibinfo {title} {Supply chain logistics with quantum and classical annealing algorithms},\ }\href@noop {} {\bibfield  {journal} {\bibinfo  {journal} {Sci. Rep.}\ }\textbf {\bibinfo {volume} {13}},\ \bibinfo {pages} {4770} (\bibinfo {year} {2023})}\BibitemShut {NoStop}%
\bibitem [{\citenamefont {Or\'us}\ \emph {et~al.}(2019)\citenamefont {Or\'us}, \citenamefont {Mugel},\ and\ \citenamefont {Lizaso}}]{Orus2019Forecasting}%
  \BibitemOpen
  \bibfield  {author} {\bibinfo {author} {\bibfnamefont {R.}~\bibnamefont {Or\'us}}, \bibinfo {author} {\bibfnamefont {S.}~\bibnamefont {Mugel}},\ and\ \bibinfo {author} {\bibfnamefont {E.}~\bibnamefont {Lizaso}},\ }\bibfield  {title} {\bibinfo {title} {Forecasting financial crashes with quantum computing},\ }\href@noop {} {\bibfield  {journal} {\bibinfo  {journal} {Phys. Rev. A}\ }\textbf {\bibinfo {volume} {99}},\ \bibinfo {pages} {060301} (\bibinfo {year} {2019})}\BibitemShut {NoStop}%
\bibitem [{\citenamefont {Mugel}\ \emph {et~al.}(2021)\citenamefont {Mugel}, \citenamefont {Abad}, \citenamefont {Bermejo}, \citenamefont {S{\'a}nchez}, \citenamefont {Lizaso},\ and\ \citenamefont {Or{\'u}s}}]{Mugel2021Hybrid}%
  \BibitemOpen
  \bibfield  {author} {\bibinfo {author} {\bibfnamefont {S.}~\bibnamefont {Mugel}}, \bibinfo {author} {\bibfnamefont {M.}~\bibnamefont {Abad}}, \bibinfo {author} {\bibfnamefont {M.}~\bibnamefont {Bermejo}}, \bibinfo {author} {\bibfnamefont {J.}~\bibnamefont {S{\'a}nchez}}, \bibinfo {author} {\bibfnamefont {E.}~\bibnamefont {Lizaso}},\ and\ \bibinfo {author} {\bibfnamefont {R.}~\bibnamefont {Or{\'u}s}},\ }\bibfield  {title} {\bibinfo {title} {Hybrid quantum investment optimization with minimal holding period},\ }\href@noop {} {\bibfield  {journal} {\bibinfo  {journal} {Sci. Rep.}\ }\textbf {\bibinfo {volume} {11}},\ \bibinfo {pages} {19587} (\bibinfo {year} {2021})}\BibitemShut {NoStop}%
\bibitem [{\citenamefont {Li}\ \emph {et~al.}(2018)\citenamefont {Li}, \citenamefont {Di~Felice}, \citenamefont {Rohs},\ and\ \citenamefont {Lidar}}]{Li2018QuantumII}%
  \BibitemOpen
  \bibfield  {author} {\bibinfo {author} {\bibfnamefont {R.~Y.}\ \bibnamefont {Li}}, \bibinfo {author} {\bibfnamefont {R.}~\bibnamefont {Di~Felice}}, \bibinfo {author} {\bibfnamefont {R.}~\bibnamefont {Rohs}},\ and\ \bibinfo {author} {\bibfnamefont {D.~A.}\ \bibnamefont {Lidar}},\ }\bibfield  {title} {\bibinfo {title} {Quantum annealing versus classical machine learning applied to a simplified computational biology problem},\ }\href@noop {} {\bibfield  {journal} {\bibinfo  {journal} {npj Quantum Inf.}\ }\textbf {\bibinfo {volume} {4}},\ \bibinfo {pages} {14} (\bibinfo {year} {2018})}\BibitemShut {NoStop}%
\bibitem [{\citenamefont {Boev}\ \emph {et~al.}(2021)\citenamefont {Boev}, \citenamefont {Rakitko}, \citenamefont {Usmanov}, \citenamefont {Kobzeva}, \citenamefont {Popov}, \citenamefont {Ilinsky}, \citenamefont {Kiktenko},\ and\ \citenamefont {Fedorov}}]{Boev2021Genome}%
  \BibitemOpen
  \bibfield  {author} {\bibinfo {author} {\bibfnamefont {A.~S.}\ \bibnamefont {Boev}}, \bibinfo {author} {\bibfnamefont {A.~S.}\ \bibnamefont {Rakitko}}, \bibinfo {author} {\bibfnamefont {S.~R.}\ \bibnamefont {Usmanov}}, \bibinfo {author} {\bibfnamefont {A.~N.}\ \bibnamefont {Kobzeva}}, \bibinfo {author} {\bibfnamefont {I.~V.}\ \bibnamefont {Popov}}, \bibinfo {author} {\bibfnamefont {V.~V.}\ \bibnamefont {Ilinsky}}, \bibinfo {author} {\bibfnamefont {E.~O.}\ \bibnamefont {Kiktenko}},\ and\ \bibinfo {author} {\bibfnamefont {A.~K.}\ \bibnamefont {Fedorov}},\ }\bibfield  {title} {\bibinfo {title} {Genome assembly using quantum and quantum-inspired annealing},\ }\href@noop {} {\bibfield  {journal} {\bibinfo  {journal} {Sci. Rep.}\ }\textbf {\bibinfo {volume} {11}},\ \bibinfo {pages} {13183} (\bibinfo {year} {2021})}\BibitemShut {NoStop}%
\bibitem [{\citenamefont {Irb\"ack}\ \emph {et~al.}(2022)\citenamefont {Irb\"ack}, \citenamefont {Knuthson}, \citenamefont {Mohanty},\ and\ \citenamefont {Peterson}}]{Irback2022Folding}%
  \BibitemOpen
  \bibfield  {author} {\bibinfo {author} {\bibfnamefont {A.}~\bibnamefont {Irb\"ack}}, \bibinfo {author} {\bibfnamefont {L.}~\bibnamefont {Knuthson}}, \bibinfo {author} {\bibfnamefont {S.}~\bibnamefont {Mohanty}},\ and\ \bibinfo {author} {\bibfnamefont {C.}~\bibnamefont {Peterson}},\ }\bibfield  {title} {\bibinfo {title} {Folding lattice proteins with quantum annealing},\ }\href@noop {} {\bibfield  {journal} {\bibinfo  {journal} {Phys. Rev. Res.}\ }\textbf {\bibinfo {volume} {4}},\ \bibinfo {pages} {043013} (\bibinfo {year} {2022})}\BibitemShut {NoStop}%
\bibitem [{\citenamefont {Gircha}\ \emph {et~al.}(2023)\citenamefont {Gircha}, \citenamefont {Boev}, \citenamefont {Avchaciov}, \citenamefont {Fedichev},\ and\ \citenamefont {Fedorov}}]{Gircha2023Hybrid}%
  \BibitemOpen
  \bibfield  {author} {\bibinfo {author} {\bibfnamefont {A.~I.}\ \bibnamefont {Gircha}}, \bibinfo {author} {\bibfnamefont {A.~S.}\ \bibnamefont {Boev}}, \bibinfo {author} {\bibfnamefont {K.}~\bibnamefont {Avchaciov}}, \bibinfo {author} {\bibfnamefont {P.~O.}\ \bibnamefont {Fedichev}},\ and\ \bibinfo {author} {\bibfnamefont {A.~K.}\ \bibnamefont {Fedorov}},\ }\bibfield  {title} {\bibinfo {title} {Hybrid quantum-classical machine learning for generative chemistry and drug design},\ }\href@noop {} {\bibfield  {journal} {\bibinfo  {journal} {Sci. Rep.}\ }\textbf {\bibinfo {volume} {13}},\ \bibinfo {pages} {8250} (\bibinfo {year} {2023})}\BibitemShut {NoStop}%
\bibitem [{\citenamefont {Messiah}(1962)}]{Messiah1962}%
  \BibitemOpen
  \bibfield  {author} {\bibinfo {author} {\bibfnamefont {A.}~\bibnamefont {Messiah}},\ }\href@noop {} {\emph {\bibinfo {title} {Quantum Mechanics, Vol. II}}}\ (\bibinfo  {publisher} {North-Holland Publishing Company},\ \bibinfo {year} {1962})\BibitemShut {NoStop}%
\bibitem [{\citenamefont {Jansen}\ \emph {et~al.}(2007)\citenamefont {Jansen}, \citenamefont {Ruskai},\ and\ \citenamefont {Seiler}}]{Jansen2007Bounds}%
  \BibitemOpen
  \bibfield  {author} {\bibinfo {author} {\bibfnamefont {S.}~\bibnamefont {Jansen}}, \bibinfo {author} {\bibfnamefont {M.-B.}\ \bibnamefont {Ruskai}},\ and\ \bibinfo {author} {\bibfnamefont {R.}~\bibnamefont {Seiler}},\ }\bibfield  {title} {\bibinfo {title} {Bounds for the adiabatic approximation with applications to quantum computation},\ }\href@noop {} {\bibfield  {journal} {\bibinfo  {journal} {J. Math. Phys.}\ }\textbf {\bibinfo {volume} {48}},\ \bibinfo {pages} {102111} (\bibinfo {year} {2007})}\BibitemShut {NoStop}%
\bibitem [{\citenamefont {J\"org}\ \emph {et~al.}(2008)\citenamefont {J\"org}, \citenamefont {Krzakala}, \citenamefont {Kurchan},\ and\ \citenamefont {Maggs}}]{Jorg2008Simple}%
  \BibitemOpen
  \bibfield  {author} {\bibinfo {author} {\bibfnamefont {T.}~\bibnamefont {J\"org}}, \bibinfo {author} {\bibfnamefont {F.}~\bibnamefont {Krzakala}}, \bibinfo {author} {\bibfnamefont {J.}~\bibnamefont {Kurchan}},\ and\ \bibinfo {author} {\bibfnamefont {A.~C.}\ \bibnamefont {Maggs}},\ }\bibfield  {title} {\bibinfo {title} {Simple glass models and their quantum annealing},\ }\href@noop {} {\bibfield  {journal} {\bibinfo  {journal} {Phys. Rev. Lett.}\ }\textbf {\bibinfo {volume} {101}},\ \bibinfo {pages} {147204} (\bibinfo {year} {2008})}\BibitemShut {NoStop}%
\bibitem [{\citenamefont {Altshuler}\ \emph {et~al.}(2010)\citenamefont {Altshuler}, \citenamefont {Krovi},\ and\ \citenamefont {Roland}}]{Altshuler2010Anderson}%
  \BibitemOpen
  \bibfield  {author} {\bibinfo {author} {\bibfnamefont {B.}~\bibnamefont {Altshuler}}, \bibinfo {author} {\bibfnamefont {H.}~\bibnamefont {Krovi}},\ and\ \bibinfo {author} {\bibfnamefont {J.}~\bibnamefont {Roland}},\ }\bibfield  {title} {\bibinfo {title} {Anderson localization makes adiabatic quantum optimization fail},\ }\href@noop {} {\bibfield  {journal} {\bibinfo  {journal} {Proc. Natl. Acad. Sci. U.S.A.}\ }\textbf {\bibinfo {volume} {107}},\ \bibinfo {pages} {12446} (\bibinfo {year} {2010})}\BibitemShut {NoStop}%
\bibitem [{\citenamefont {Foini}\ \emph {et~al.}(2010)\citenamefont {Foini}, \citenamefont {Semerjian},\ and\ \citenamefont {Zamponi}}]{Foini2010Solvable}%
  \BibitemOpen
  \bibfield  {author} {\bibinfo {author} {\bibfnamefont {L.}~\bibnamefont {Foini}}, \bibinfo {author} {\bibfnamefont {G.}~\bibnamefont {Semerjian}},\ and\ \bibinfo {author} {\bibfnamefont {F.}~\bibnamefont {Zamponi}},\ }\bibfield  {title} {\bibinfo {title} {Solvable model of quantum random optimization problems},\ }\href@noop {} {\bibfield  {journal} {\bibinfo  {journal} {Phys. Rev. Lett.}\ }\textbf {\bibinfo {volume} {105}},\ \bibinfo {pages} {167204} (\bibinfo {year} {2010})}\BibitemShut {NoStop}%
\bibitem [{\citenamefont {Bapst}\ \emph {et~al.}(2013)\citenamefont {Bapst}, \citenamefont {Foini}, \citenamefont {Krzakala}, \citenamefont {Semerjian},\ and\ \citenamefont {Zamponi}}]{Bapst2013Quantum}%
  \BibitemOpen
  \bibfield  {author} {\bibinfo {author} {\bibfnamefont {V.}~\bibnamefont {Bapst}}, \bibinfo {author} {\bibfnamefont {L.}~\bibnamefont {Foini}}, \bibinfo {author} {\bibfnamefont {F.}~\bibnamefont {Krzakala}}, \bibinfo {author} {\bibfnamefont {G.}~\bibnamefont {Semerjian}},\ and\ \bibinfo {author} {\bibfnamefont {F.}~\bibnamefont {Zamponi}},\ }\bibfield  {title} {\bibinfo {title} {The quantum adiabatic algorithm applied to random optimization problems: {T}he quantum spin glass perspective},\ }\href@noop {} {\bibfield  {journal} {\bibinfo  {journal} {Phys. Rep.}\ }\textbf {\bibinfo {volume} {523}},\ \bibinfo {pages} {127} (\bibinfo {year} {2013})}\BibitemShut {NoStop}%
\bibitem [{\citenamefont {Knysh}(2016)}]{Knysh2016ZeroTemperature}%
  \BibitemOpen
  \bibfield  {author} {\bibinfo {author} {\bibfnamefont {S.}~\bibnamefont {Knysh}},\ }\bibfield  {title} {\bibinfo {title} {Zero-temperature quantum annealing bottlenecks in the spin-glass phase},\ }\href@noop {} {\bibfield  {journal} {\bibinfo  {journal} {Nat. Commun.}\ }\textbf {\bibinfo {volume} {7}},\ \bibinfo {pages} {12370} (\bibinfo {year} {2016})}\BibitemShut {NoStop}%
\bibitem [{\citenamefont {Baldwin}\ and\ \citenamefont {Laumann}(2018)}]{Baldwin2018Quantum}%
  \BibitemOpen
  \bibfield  {author} {\bibinfo {author} {\bibfnamefont {C.~L.}\ \bibnamefont {Baldwin}}\ and\ \bibinfo {author} {\bibfnamefont {C.~R.}\ \bibnamefont {Laumann}},\ }\bibfield  {title} {\bibinfo {title} {Quantum algorithm for energy matching in hard optimization problems},\ }\href@noop {} {\bibfield  {journal} {\bibinfo  {journal} {Phys. Rev. B}\ }\textbf {\bibinfo {volume} {97}},\ \bibinfo {pages} {224201} (\bibinfo {year} {2018})}\BibitemShut {NoStop}%
\bibitem [{\citenamefont {King}\ \emph {et~al.}(2022)\citenamefont {King}, \citenamefont {Suzuki}, \citenamefont {Raymond}, \citenamefont {Zucca}, \citenamefont {Lanting}, \citenamefont {Altomare}, \citenamefont {Berkley}, \citenamefont {Ejtemaee}, \citenamefont {Hoskinson}, \citenamefont {Huang}, \citenamefont {Ladizinsky}, \citenamefont {MacDonald}, \citenamefont {Marsden}, \citenamefont {Oh}, \citenamefont {Poulin-Lamarre}, \citenamefont {Reis}, \citenamefont {Rich}, \citenamefont {Sato}, \citenamefont {Whittaker}, \citenamefont {Yao}, \citenamefont {Harris}, \citenamefont {Lidar}, \citenamefont {Nishimori},\ and\ \citenamefont {Amin}}]{King2022Coherent}%
  \BibitemOpen
  \bibfield  {author} {\bibinfo {author} {\bibfnamefont {A.~D.}\ \bibnamefont {King}}, \bibinfo {author} {\bibfnamefont {S.}~\bibnamefont {Suzuki}}, \bibinfo {author} {\bibfnamefont {J.}~\bibnamefont {Raymond}}, \bibinfo {author} {\bibfnamefont {A.}~\bibnamefont {Zucca}}, \bibinfo {author} {\bibfnamefont {T.}~\bibnamefont {Lanting}}, \bibinfo {author} {\bibfnamefont {F.}~\bibnamefont {Altomare}}, \bibinfo {author} {\bibfnamefont {A.~J.}\ \bibnamefont {Berkley}}, \bibinfo {author} {\bibfnamefont {S.}~\bibnamefont {Ejtemaee}}, \bibinfo {author} {\bibfnamefont {E.}~\bibnamefont {Hoskinson}}, \bibinfo {author} {\bibfnamefont {S.}~\bibnamefont {Huang}}, \bibinfo {author} {\bibfnamefont {E.}~\bibnamefont {Ladizinsky}}, \bibinfo {author} {\bibfnamefont {A.~J.~R.}\ \bibnamefont {MacDonald}}, \bibinfo {author} {\bibfnamefont {G.}~\bibnamefont {Marsden}}, \bibinfo {author} {\bibfnamefont {T.}~\bibnamefont {Oh}}, \bibinfo {author} {\bibfnamefont {G.}~\bibnamefont {Poulin-Lamarre}}, \bibinfo {author} {\bibfnamefont
  {M.}~\bibnamefont {Reis}}, \bibinfo {author} {\bibfnamefont {C.}~\bibnamefont {Rich}}, \bibinfo {author} {\bibfnamefont {Y.}~\bibnamefont {Sato}}, \bibinfo {author} {\bibfnamefont {J.~D.}\ \bibnamefont {Whittaker}}, \bibinfo {author} {\bibfnamefont {J.}~\bibnamefont {Yao}}, \bibinfo {author} {\bibfnamefont {R.}~\bibnamefont {Harris}}, \bibinfo {author} {\bibfnamefont {D.~A.}\ \bibnamefont {Lidar}}, \bibinfo {author} {\bibfnamefont {H.}~\bibnamefont {Nishimori}},\ and\ \bibinfo {author} {\bibfnamefont {M.~H.}\ \bibnamefont {Amin}},\ }\bibfield  {title} {\bibinfo {title} {Coherent quantum annealing in a programmable 2,000 qubit {I}sing chain},\ }\href@noop {} {\bibfield  {journal} {\bibinfo  {journal} {Nat. Phys.}\ }\textbf {\bibinfo {volume} {18}},\ \bibinfo {pages} {1324} (\bibinfo {year} {2022})}\BibitemShut {NoStop}%
\bibitem [{\citenamefont {King}\ \emph {et~al.}(2023)\citenamefont {King}, \citenamefont {Raymond}, \citenamefont {Lanting}, \citenamefont {Harris}, \citenamefont {Zucca}, \citenamefont {Altomare}, \citenamefont {Berkley}, \citenamefont {Boothby}, \citenamefont {Ejtemaee}, \citenamefont {Enderud}, \citenamefont {Hoskinson}, \citenamefont {Huang}, \citenamefont {Ladizinsky}, \citenamefont {MacDonald}, \citenamefont {Marsden}, \citenamefont {Molavi}, \citenamefont {Oh}, \citenamefont {Poulin-Lamarre}, \citenamefont {Reis}, \citenamefont {Rich}, \citenamefont {Sato}, \citenamefont {Tsai}, \citenamefont {Volkmann}, \citenamefont {Whittaker}, \citenamefont {Yao}, \citenamefont {Sandvik},\ and\ \citenamefont {Amin}}]{King2023Quantum}%
  \BibitemOpen
  \bibfield  {author} {\bibinfo {author} {\bibfnamefont {A.~D.}\ \bibnamefont {King}}, \bibinfo {author} {\bibfnamefont {J.}~\bibnamefont {Raymond}}, \bibinfo {author} {\bibfnamefont {T.}~\bibnamefont {Lanting}}, \bibinfo {author} {\bibfnamefont {R.}~\bibnamefont {Harris}}, \bibinfo {author} {\bibfnamefont {A.}~\bibnamefont {Zucca}}, \bibinfo {author} {\bibfnamefont {F.}~\bibnamefont {Altomare}}, \bibinfo {author} {\bibfnamefont {A.~J.}\ \bibnamefont {Berkley}}, \bibinfo {author} {\bibfnamefont {K.}~\bibnamefont {Boothby}}, \bibinfo {author} {\bibfnamefont {S.}~\bibnamefont {Ejtemaee}}, \bibinfo {author} {\bibfnamefont {C.}~\bibnamefont {Enderud}}, \bibinfo {author} {\bibfnamefont {E.}~\bibnamefont {Hoskinson}}, \bibinfo {author} {\bibfnamefont {S.}~\bibnamefont {Huang}}, \bibinfo {author} {\bibfnamefont {E.}~\bibnamefont {Ladizinsky}}, \bibinfo {author} {\bibfnamefont {A.~J.~R.}\ \bibnamefont {MacDonald}}, \bibinfo {author} {\bibfnamefont {G.}~\bibnamefont {Marsden}}, \bibinfo {author} {\bibfnamefont
  {R.}~\bibnamefont {Molavi}}, \bibinfo {author} {\bibfnamefont {T.}~\bibnamefont {Oh}}, \bibinfo {author} {\bibfnamefont {G.}~\bibnamefont {Poulin-Lamarre}}, \bibinfo {author} {\bibfnamefont {M.}~\bibnamefont {Reis}}, \bibinfo {author} {\bibfnamefont {C.}~\bibnamefont {Rich}}, \bibinfo {author} {\bibfnamefont {Y.}~\bibnamefont {Sato}}, \bibinfo {author} {\bibfnamefont {N.}~\bibnamefont {Tsai}}, \bibinfo {author} {\bibfnamefont {M.}~\bibnamefont {Volkmann}}, \bibinfo {author} {\bibfnamefont {J.~D.}\ \bibnamefont {Whittaker}}, \bibinfo {author} {\bibfnamefont {J.}~\bibnamefont {Yao}}, \bibinfo {author} {\bibfnamefont {A.~W.}\ \bibnamefont {Sandvik}},\ and\ \bibinfo {author} {\bibfnamefont {M.~H.}\ \bibnamefont {Amin}},\ }\bibfield  {title} {\bibinfo {title} {Quantum critical dynamics in a 5,000-qubit programmable spin glass},\ }\href@noop {} {\bibfield  {journal} {\bibinfo  {journal} {Nature}\ }\textbf {\bibinfo {volume} {617}},\ \bibinfo {pages} {61} (\bibinfo {year} {2023})}\BibitemShut {NoStop}%
\bibitem [{\citenamefont {King}\ \emph {et~al.}(2025)\citenamefont {King}, \citenamefont {Nocera}, \citenamefont {Rams}, \citenamefont {Dziarmaga}, \citenamefont {Wiersema}, \citenamefont {Bernoudy}, \citenamefont {Raymond}, \citenamefont {Kaushal}, \citenamefont {Heinsdorf}, \citenamefont {Harris}, \citenamefont {Boothby}, \citenamefont {Altomare}, \citenamefont {Asad}, \citenamefont {Berkley}, \citenamefont {Boschnak}, \citenamefont {Chern}, \citenamefont {Christiani}, \citenamefont {Cibere}, \citenamefont {Connor}, \citenamefont {Dehn}, \citenamefont {Deshpande}, \citenamefont {Ejtemaee}, \citenamefont {Farre}, \citenamefont {Hamer}, \citenamefont {Hoskinson}, \citenamefont {Huang}, \citenamefont {Johnson}, \citenamefont {Kortas}, \citenamefont {Ladizinsky}, \citenamefont {Lanting}, \citenamefont {Lai}, \citenamefont {Li}, \citenamefont {MacDonald}, \citenamefont {Marsden}, \citenamefont {McGeoch}, \citenamefont {Molavi}, \citenamefont {Oh}, \citenamefont {Neufeld}, \citenamefont {Norouzpour},
  \citenamefont {Pasvolsky}, \citenamefont {Poitras}, \citenamefont {Poulin-Lamarre}, \citenamefont {Prescott}, \citenamefont {Reis}, \citenamefont {Rich}, \citenamefont {Samani}, \citenamefont {Sheldan}, \citenamefont {Smirnov}, \citenamefont {Sterpka}, \citenamefont {Trullas~Clavera}, \citenamefont {Tsai}, \citenamefont {Volkmann}, \citenamefont {Whiticar}, \citenamefont {Whittaker}, \citenamefont {Wilkinson}, \citenamefont {Yao}, \citenamefont {Yi}, \citenamefont {Sandvik}, \citenamefont {Alvarez}, \citenamefont {Melko}, \citenamefont {Carrasquilla}, \citenamefont {Franz},\ and\ \citenamefont {Amin}}]{King2025BeyondClassical}%
  \BibitemOpen
  \bibfield  {author} {\bibinfo {author} {\bibfnamefont {A.~D.}\ \bibnamefont {King}}, \bibinfo {author} {\bibfnamefont {A.}~\bibnamefont {Nocera}}, \bibinfo {author} {\bibfnamefont {M.~M.}\ \bibnamefont {Rams}}, \bibinfo {author} {\bibfnamefont {J.}~\bibnamefont {Dziarmaga}}, \bibinfo {author} {\bibfnamefont {R.}~\bibnamefont {Wiersema}}, \bibinfo {author} {\bibfnamefont {W.}~\bibnamefont {Bernoudy}}, \bibinfo {author} {\bibfnamefont {J.}~\bibnamefont {Raymond}}, \bibinfo {author} {\bibfnamefont {N.}~\bibnamefont {Kaushal}}, \bibinfo {author} {\bibfnamefont {N.}~\bibnamefont {Heinsdorf}}, \bibinfo {author} {\bibfnamefont {R.}~\bibnamefont {Harris}}, \bibinfo {author} {\bibfnamefont {K.}~\bibnamefont {Boothby}}, \bibinfo {author} {\bibfnamefont {F.}~\bibnamefont {Altomare}}, \bibinfo {author} {\bibfnamefont {M.}~\bibnamefont {Asad}}, \bibinfo {author} {\bibfnamefont {A.~J.}\ \bibnamefont {Berkley}}, \bibinfo {author} {\bibfnamefont {M.}~\bibnamefont {Boschnak}}, \bibinfo {author} {\bibfnamefont
  {K.}~\bibnamefont {Chern}}, \bibinfo {author} {\bibfnamefont {H.}~\bibnamefont {Christiani}}, \bibinfo {author} {\bibfnamefont {S.}~\bibnamefont {Cibere}}, \bibinfo {author} {\bibfnamefont {J.}~\bibnamefont {Connor}}, \bibinfo {author} {\bibfnamefont {M.~H.}\ \bibnamefont {Dehn}}, \bibinfo {author} {\bibfnamefont {R.}~\bibnamefont {Deshpande}}, \bibinfo {author} {\bibfnamefont {S.}~\bibnamefont {Ejtemaee}}, \bibinfo {author} {\bibfnamefont {P.}~\bibnamefont {Farre}}, \bibinfo {author} {\bibfnamefont {K.}~\bibnamefont {Hamer}}, \bibinfo {author} {\bibfnamefont {E.}~\bibnamefont {Hoskinson}}, \bibinfo {author} {\bibfnamefont {S.}~\bibnamefont {Huang}}, \bibinfo {author} {\bibfnamefont {M.~W.}\ \bibnamefont {Johnson}}, \bibinfo {author} {\bibfnamefont {S.}~\bibnamefont {Kortas}}, \bibinfo {author} {\bibfnamefont {E.}~\bibnamefont {Ladizinsky}}, \bibinfo {author} {\bibfnamefont {T.}~\bibnamefont {Lanting}}, \bibinfo {author} {\bibfnamefont {T.}~\bibnamefont {Lai}}, \bibinfo {author} {\bibfnamefont
  {R.}~\bibnamefont {Li}}, \bibinfo {author} {\bibfnamefont {A.~J.~R.}\ \bibnamefont {MacDonald}}, \bibinfo {author} {\bibfnamefont {G.}~\bibnamefont {Marsden}}, \bibinfo {author} {\bibfnamefont {C.~C.}\ \bibnamefont {McGeoch}}, \bibinfo {author} {\bibfnamefont {R.}~\bibnamefont {Molavi}}, \bibinfo {author} {\bibfnamefont {T.}~\bibnamefont {Oh}}, \bibinfo {author} {\bibfnamefont {R.}~\bibnamefont {Neufeld}}, \bibinfo {author} {\bibfnamefont {M.}~\bibnamefont {Norouzpour}}, \bibinfo {author} {\bibfnamefont {J.}~\bibnamefont {Pasvolsky}}, \bibinfo {author} {\bibfnamefont {P.}~\bibnamefont {Poitras}}, \bibinfo {author} {\bibfnamefont {G.}~\bibnamefont {Poulin-Lamarre}}, \bibinfo {author} {\bibfnamefont {T.}~\bibnamefont {Prescott}}, \bibinfo {author} {\bibfnamefont {M.}~\bibnamefont {Reis}}, \bibinfo {author} {\bibfnamefont {C.}~\bibnamefont {Rich}}, \bibinfo {author} {\bibfnamefont {M.}~\bibnamefont {Samani}}, \bibinfo {author} {\bibfnamefont {B.}~\bibnamefont {Sheldan}}, \bibinfo {author} {\bibfnamefont
  {A.}~\bibnamefont {Smirnov}}, \bibinfo {author} {\bibfnamefont {E.}~\bibnamefont {Sterpka}}, \bibinfo {author} {\bibfnamefont {B.}~\bibnamefont {Trullas~Clavera}}, \bibinfo {author} {\bibfnamefont {N.}~\bibnamefont {Tsai}}, \bibinfo {author} {\bibfnamefont {M.}~\bibnamefont {Volkmann}}, \bibinfo {author} {\bibfnamefont {A.~M.}\ \bibnamefont {Whiticar}}, \bibinfo {author} {\bibfnamefont {J.~D.}\ \bibnamefont {Whittaker}}, \bibinfo {author} {\bibfnamefont {W.}~\bibnamefont {Wilkinson}}, \bibinfo {author} {\bibfnamefont {J.}~\bibnamefont {Yao}}, \bibinfo {author} {\bibfnamefont {T.~J.}\ \bibnamefont {Yi}}, \bibinfo {author} {\bibfnamefont {A.~W.}\ \bibnamefont {Sandvik}}, \bibinfo {author} {\bibfnamefont {G.}~\bibnamefont {Alvarez}}, \bibinfo {author} {\bibfnamefont {R.~G.}\ \bibnamefont {Melko}}, \bibinfo {author} {\bibfnamefont {J.}~\bibnamefont {Carrasquilla}}, \bibinfo {author} {\bibfnamefont {M.}~\bibnamefont {Franz}},\ and\ \bibinfo {author} {\bibfnamefont {M.~H.}\ \bibnamefont {Amin}},\ }\bibfield
  {title} {\bibinfo {title} {Beyond-classical computation in quantum simulation},\ }\href@noop {} {\bibfield  {journal} {\bibinfo  {journal} {Science}\ }\textbf {\bibinfo {volume} {388}},\ \bibinfo {pages} {199} (\bibinfo {year} {2025})}\BibitemShut {NoStop}%
\bibitem [{\citenamefont {Braida}\ \emph {et~al.}(2024)\citenamefont {Braida}, \citenamefont {Martiel},\ and\ \citenamefont {Todinca}}]{Braida2024Tight}%
  \BibitemOpen
  \bibfield  {author} {\bibinfo {author} {\bibfnamefont {A.}~\bibnamefont {Braida}}, \bibinfo {author} {\bibfnamefont {S.}~\bibnamefont {Martiel}},\ and\ \bibinfo {author} {\bibfnamefont {I.}~\bibnamefont {Todinca}},\ }\bibfield  {title} {\bibinfo {title} {Tight {L}ieb-{R}obinson bound for approximation ratio in quantum annealing},\ }\href@noop {} {\bibfield  {journal} {\bibinfo  {journal} {npj Quantum Inf.}\ }\textbf {\bibinfo {volume} {10}},\ \bibinfo {pages} {40} (\bibinfo {year} {2024})}\BibitemShut {NoStop}%
\bibitem [{\citenamefont {Zhang}\ \emph {et~al.}(2024)\citenamefont {Zhang}, \citenamefont {Boothby},\ and\ \citenamefont {Kamenev}}]{Zhang2024Cyclic}%
  \BibitemOpen
  \bibfield  {author} {\bibinfo {author} {\bibfnamefont {H.}~\bibnamefont {Zhang}}, \bibinfo {author} {\bibfnamefont {K.}~\bibnamefont {Boothby}},\ and\ \bibinfo {author} {\bibfnamefont {A.}~\bibnamefont {Kamenev}},\ }\bibfield  {title} {\bibinfo {title} {Cyclic quantum annealing: {S}earching for deep low-energy states in 5000-qubit spin glass},\ }\href@noop {} {\bibfield  {journal} {\bibinfo  {journal} {Sci. Rep.}\ }\textbf {\bibinfo {volume} {14}},\ \bibinfo {pages} {30784} (\bibinfo {year} {2024})}\BibitemShut {NoStop}%
\bibitem [{\citenamefont {Munoz-Bauza}\ and\ \citenamefont {Lidar}(2025)}]{MunozBauza2025Scaling}%
  \BibitemOpen
  \bibfield  {author} {\bibinfo {author} {\bibfnamefont {H.}~\bibnamefont {Munoz-Bauza}}\ and\ \bibinfo {author} {\bibfnamefont {D.}~\bibnamefont {Lidar}},\ }\bibfield  {title} {\bibinfo {title} {Scaling advantage in approximate optimization with quantum annealing},\ }\href@noop {} {\bibfield  {journal} {\bibinfo  {journal} {Phys. Rev. Lett.}\ }\textbf {\bibinfo {volume} {134}},\ \bibinfo {pages} {160601} (\bibinfo {year} {2025})}\BibitemShut {NoStop}%
\bibitem [{\citenamefont {Crisanti}\ and\ \citenamefont {Sommers}(1992)}]{Crisanti1992Spherical}%
  \BibitemOpen
  \bibfield  {author} {\bibinfo {author} {\bibfnamefont {A.}~\bibnamefont {Crisanti}}\ and\ \bibinfo {author} {\bibfnamefont {H.~J.}\ \bibnamefont {Sommers}},\ }\bibfield  {title} {\bibinfo {title} {The spherical p-spin interaction spin glass model: {T}he statics},\ }\href@noop {} {\bibfield  {journal} {\bibinfo  {journal} {Z. Phys. B: Condens. Matter}\ }\textbf {\bibinfo {volume} {87}},\ \bibinfo {pages} {341} (\bibinfo {year} {1992})}\BibitemShut {NoStop}%
\bibitem [{\citenamefont {Crisanti}\ \emph {et~al.}(1993)\citenamefont {Crisanti}, \citenamefont {Horner},\ and\ \citenamefont {Sommers}}]{Crisanti1993Spherical}%
  \BibitemOpen
  \bibfield  {author} {\bibinfo {author} {\bibfnamefont {A.}~\bibnamefont {Crisanti}}, \bibinfo {author} {\bibfnamefont {H.}~\bibnamefont {Horner}},\ and\ \bibinfo {author} {\bibfnamefont {H.~J.}\ \bibnamefont {Sommers}},\ }\bibfield  {title} {\bibinfo {title} {The spherical p-spin interaction spin-glass model: {T}he dynamics},\ }\href@noop {} {\bibfield  {journal} {\bibinfo  {journal} {Z. Phys. B: Condens. Matter}\ }\textbf {\bibinfo {volume} {92}},\ \bibinfo {pages} {257} (\bibinfo {year} {1993})}\BibitemShut {NoStop}%
\bibitem [{\citenamefont {Castellani}\ and\ \citenamefont {Cavagna}(2005)}]{Castellani2005Spin}%
  \BibitemOpen
  \bibfield  {author} {\bibinfo {author} {\bibfnamefont {T.}~\bibnamefont {Castellani}}\ and\ \bibinfo {author} {\bibfnamefont {A.}~\bibnamefont {Cavagna}},\ }\bibfield  {title} {\bibinfo {title} {Spin-glass theory for pedestrians},\ }\href@noop {} {\bibfield  {journal} {\bibinfo  {journal} {J. Stat. Mech.: Theory Exp.}\ }\textbf {\bibinfo {volume} {2005}},\ \bibinfo {pages} {P05012}}\BibitemShut {NoStop}%
\bibitem [{\citenamefont {Mezard}\ \emph {et~al.}(1987)\citenamefont {Mezard}, \citenamefont {Parisi},\ and\ \citenamefont {Virasoro}}]{Mezard1987}%
  \BibitemOpen
  \bibfield  {author} {\bibinfo {author} {\bibfnamefont {M.}~\bibnamefont {Mezard}}, \bibinfo {author} {\bibfnamefont {G.}~\bibnamefont {Parisi}},\ and\ \bibinfo {author} {\bibfnamefont {M.~A.}\ \bibnamefont {Virasoro}},\ }\href@noop {} {\emph {\bibinfo {title} {Spin Glass Theory and Beyond}}}\ (\bibinfo  {publisher} {World Scientific},\ \bibinfo {year} {1987})\BibitemShut {NoStop}%
\bibitem [{\citenamefont {Fischer}\ and\ \citenamefont {Hertz}(1991)}]{Fischer1991}%
  \BibitemOpen
  \bibfield  {author} {\bibinfo {author} {\bibfnamefont {K.~H.}\ \bibnamefont {Fischer}}\ and\ \bibinfo {author} {\bibfnamefont {J.~A.}\ \bibnamefont {Hertz}},\ }\href@noop {} {\emph {\bibinfo {title} {Spin Glasses}}}\ (\bibinfo  {publisher} {Cambridge University Press},\ \bibinfo {year} {1991})\BibitemShut {NoStop}%
\bibitem [{\citenamefont {Nishimori}(2001)}]{Nishimori2001}%
  \BibitemOpen
  \bibfield  {author} {\bibinfo {author} {\bibfnamefont {H.}~\bibnamefont {Nishimori}},\ }\href@noop {} {\emph {\bibinfo {title} {Statistical Physics of Spin Glasses and Information Processing}}}\ (\bibinfo  {publisher} {Clarendon Press},\ \bibinfo {year} {2001})\BibitemShut {NoStop}%
\bibitem [{\citenamefont {Mezard}\ and\ \citenamefont {Montanari}(2009)}]{Mezard2009}%
  \BibitemOpen
  \bibfield  {author} {\bibinfo {author} {\bibfnamefont {M.}~\bibnamefont {Mezard}}\ and\ \bibinfo {author} {\bibfnamefont {A.}~\bibnamefont {Montanari}},\ }\href@noop {} {\emph {\bibinfo {title} {Information, Physics, and Computation}}}\ (\bibinfo  {publisher} {Oxford University Press},\ \bibinfo {year} {2009})\BibitemShut {NoStop}%
\bibitem [{\citenamefont {Cugliandolo}\ and\ \citenamefont {Kurchan}(1993)}]{Cugliandolo1993Analytical}%
  \BibitemOpen
  \bibfield  {author} {\bibinfo {author} {\bibfnamefont {L.~F.}\ \bibnamefont {Cugliandolo}}\ and\ \bibinfo {author} {\bibfnamefont {J.}~\bibnamefont {Kurchan}},\ }\bibfield  {title} {\bibinfo {title} {Analytical solution of the off-equilibrium dynamics of a long-range spin-glass model},\ }\href@noop {} {\bibfield  {journal} {\bibinfo  {journal} {Phys. Rev. Lett.}\ }\textbf {\bibinfo {volume} {71}},\ \bibinfo {pages} {173} (\bibinfo {year} {1993})}\BibitemShut {NoStop}%
\bibitem [{\citenamefont {Barrat}\ \emph {et~al.}(1996)\citenamefont {Barrat}, \citenamefont {Burioni},\ and\ \citenamefont {M{\'e}zard}}]{Barrat1996Dynamics}%
  \BibitemOpen
  \bibfield  {author} {\bibinfo {author} {\bibfnamefont {A.}~\bibnamefont {Barrat}}, \bibinfo {author} {\bibfnamefont {R.}~\bibnamefont {Burioni}},\ and\ \bibinfo {author} {\bibfnamefont {M.}~\bibnamefont {M{\'e}zard}},\ }\bibfield  {title} {\bibinfo {title} {Dynamics within metastable states in a mean-field spin glass},\ }\href@noop {} {\bibfield  {journal} {\bibinfo  {journal} {J. Phys. A: Math. Theor.}\ }\textbf {\bibinfo {volume} {29}},\ \bibinfo {pages} {L81} (\bibinfo {year} {1996})}\BibitemShut {NoStop}%
\bibitem [{\citenamefont {Cugliandolo}\ and\ \citenamefont {Lozano}(1999)}]{Cugliandolo1999RealTime}%
  \BibitemOpen
  \bibfield  {author} {\bibinfo {author} {\bibfnamefont {L.~F.}\ \bibnamefont {Cugliandolo}}\ and\ \bibinfo {author} {\bibfnamefont {G.}~\bibnamefont {Lozano}},\ }\bibfield  {title} {\bibinfo {title} {Real-time nonequilibrium dynamics of quantum glassy systems},\ }\href@noop {} {\bibfield  {journal} {\bibinfo  {journal} {Phys. Rev. B}\ }\textbf {\bibinfo {volume} {59}},\ \bibinfo {pages} {915} (\bibinfo {year} {1999})}\BibitemShut {NoStop}%
\bibitem [{\citenamefont {Cugliandolo}\ \emph {et~al.}(2017)\citenamefont {Cugliandolo}, \citenamefont {Lozano},\ and\ \citenamefont {Nessi}}]{Cugliandolo2017NonEquilibrium}%
  \BibitemOpen
  \bibfield  {author} {\bibinfo {author} {\bibfnamefont {L.~F.}\ \bibnamefont {Cugliandolo}}, \bibinfo {author} {\bibfnamefont {G.~S.}\ \bibnamefont {Lozano}},\ and\ \bibinfo {author} {\bibfnamefont {E.~N.}\ \bibnamefont {Nessi}},\ }\bibfield  {title} {\bibinfo {title} {Non equilibrium dynamics of isolated disordered systems: {T}he classical {H}amiltonian p-spin model},\ }\href@noop {} {\bibfield  {journal} {\bibinfo  {journal} {J. Stat. Mech.: Theory Exp.}\ }\textbf {\bibinfo {volume} {2017}},\ \bibinfo {pages} {083301}}\BibitemShut {NoStop}%
\bibitem [{\citenamefont {Crisanti}\ and\ \citenamefont {Sommers}(1995)}]{Crisanti1995ThoulessAndersonPalmer}%
  \BibitemOpen
  \bibfield  {author} {\bibinfo {author} {\bibfnamefont {A.}~\bibnamefont {Crisanti}}\ and\ \bibinfo {author} {\bibfnamefont {H.-J.}\ \bibnamefont {Sommers}},\ }\bibfield  {title} {\bibinfo {title} {Thouless-{A}nderson-{P}almer approach to the spherical p-spin spin glass model},\ }\href@noop {} {\bibfield  {journal} {\bibinfo  {journal} {J. Phys. I France}\ }\textbf {\bibinfo {volume} {5}},\ \bibinfo {pages} {805} (\bibinfo {year} {1995})}\BibitemShut {NoStop}%
\bibitem [{\citenamefont {Bray}\ and\ \citenamefont {Dean}(2007)}]{Bray2007Statistics}%
  \BibitemOpen
  \bibfield  {author} {\bibinfo {author} {\bibfnamefont {A.~J.}\ \bibnamefont {Bray}}\ and\ \bibinfo {author} {\bibfnamefont {D.~S.}\ \bibnamefont {Dean}},\ }\bibfield  {title} {\bibinfo {title} {Statistics of critical points of {G}aussian fields on large-dimensional spaces},\ }\href@noop {} {\bibfield  {journal} {\bibinfo  {journal} {Phys. Rev. Lett.}\ }\textbf {\bibinfo {volume} {98}},\ \bibinfo {pages} {150201} (\bibinfo {year} {2007})}\BibitemShut {NoStop}%
\bibitem [{ter()}]{terminology_note}%
  \BibitemOpen
  \href@noop {} {}\bibinfo {note} {For brevity, we refer to any Monte-Carlo or Langevin dynamics with time-dependent temperature as a ``simulated annealing protocol'', even if the time-dependence consists of sudden changes in temperature that would more commonly be called ``quenches''.}\BibitemShut {Stop}%
\bibitem [{\citenamefont {Folena}\ \emph {et~al.}(2020)\citenamefont {Folena}, \citenamefont {Franz},\ and\ \citenamefont {Ricci-Tersenghi}}]{Folena2020Rethinking}%
  \BibitemOpen
  \bibfield  {author} {\bibinfo {author} {\bibfnamefont {G.}~\bibnamefont {Folena}}, \bibinfo {author} {\bibfnamefont {S.}~\bibnamefont {Franz}},\ and\ \bibinfo {author} {\bibfnamefont {F.}~\bibnamefont {Ricci-Tersenghi}},\ }\bibfield  {title} {\bibinfo {title} {Rethinking mean-field glassy dynamics and its relation with the energy landscape: The surprising case of the spherical mixed p-spin model},\ }\href@noop {} {\bibfield  {journal} {\bibinfo  {journal} {Phys. Rev. X}\ }\textbf {\bibinfo {volume} {10}},\ \bibinfo {pages} {031045} (\bibinfo {year} {2020})}\BibitemShut {NoStop}%
\bibitem [{\citenamefont {Folena}\ \emph {et~al.}(2021)\citenamefont {Folena}, \citenamefont {Franz},\ and\ \citenamefont {Ricci-Tersenghi}}]{Folena2021Gradient}%
  \BibitemOpen
  \bibfield  {author} {\bibinfo {author} {\bibfnamefont {G.}~\bibnamefont {Folena}}, \bibinfo {author} {\bibfnamefont {S.}~\bibnamefont {Franz}},\ and\ \bibinfo {author} {\bibfnamefont {F.}~\bibnamefont {Ricci-Tersenghi}},\ }\bibfield  {title} {\bibinfo {title} {Gradient descent dynamics in the mixed p-spin spherical model: {F}inite-size simulations and comparison with mean-field integration},\ }\href@noop {} {\bibfield  {journal} {\bibinfo  {journal} {J. Stat. Mech.: Theory Exp.}\ }\textbf {\bibinfo {volume} {2021}},\ \bibinfo {pages} {033302}}\BibitemShut {NoStop}%
\bibitem [{tra()}]{transverse_field_note}%
  \BibitemOpen
  \href@noop {} {}\bibinfo {note} {Note that the ground state of the kinetic energy is fully delocalized in the computational ($\hat{X}$) basis, exactly analogous to a transverse field in Ising models.}\BibitemShut {Stop}%
\bibitem [{\citenamefont {Cugliandolo}\ \emph {et~al.}(2019)\citenamefont {Cugliandolo}, \citenamefont {Lozano},\ and\ \citenamefont {Nessi}}]{Cugliandolo2019Role}%
  \BibitemOpen
  \bibfield  {author} {\bibinfo {author} {\bibfnamefont {L.~F.}\ \bibnamefont {Cugliandolo}}, \bibinfo {author} {\bibfnamefont {G.~S.}\ \bibnamefont {Lozano}},\ and\ \bibinfo {author} {\bibfnamefont {N.}~\bibnamefont {Nessi}},\ }\bibfield  {title} {\bibinfo {title} {Role of initial conditions in the dynamics of quantum glassy systems},\ }\href@noop {} {\bibfield  {journal} {\bibinfo  {journal} {J. Stat. Mech.: Theory Exp.}\ }\textbf {\bibinfo {volume} {2019}},\ \bibinfo {pages} {023301}}\BibitemShut {NoStop}%
\bibitem [{\citenamefont {Thomson}\ \emph {et~al.}(2020)\citenamefont {Thomson}, \citenamefont {Urbani},\ and\ \citenamefont {Schir\'o}}]{Thomson2020Quantum}%
  \BibitemOpen
  \bibfield  {author} {\bibinfo {author} {\bibfnamefont {S.~J.}\ \bibnamefont {Thomson}}, \bibinfo {author} {\bibfnamefont {P.}~\bibnamefont {Urbani}},\ and\ \bibinfo {author} {\bibfnamefont {M.}~\bibnamefont {Schir\'o}},\ }\bibfield  {title} {\bibinfo {title} {Quantum quenches in isolated quantum glasses out of equilibrium},\ }\href@noop {} {\bibfield  {journal} {\bibinfo  {journal} {Phys. Rev. Lett.}\ }\textbf {\bibinfo {volume} {125}},\ \bibinfo {pages} {120602} (\bibinfo {year} {2020})}\BibitemShut {NoStop}%
\bibitem [{\citenamefont {Auffinger}\ and\ \citenamefont {Arous}(2013)}]{Auffinger2013Complexity}%
  \BibitemOpen
  \bibfield  {author} {\bibinfo {author} {\bibfnamefont {A.}~\bibnamefont {Auffinger}}\ and\ \bibinfo {author} {\bibfnamefont {G.~B.}\ \bibnamefont {Arous}},\ }\bibfield  {title} {\bibinfo {title} {Complexity of random smooth functions on the high-dimensional sphere},\ }\href@noop {} {\bibfield  {journal} {\bibinfo  {journal} {Ann. Probab.}\ }\textbf {\bibinfo {volume} {41}},\ \bibinfo {pages} {4214} (\bibinfo {year} {2013})}\BibitemShut {NoStop}%
\bibitem [{fit()}]{fitting_note}%
  \BibitemOpen
  \href@noop {} {}\bibinfo {note} {Since $\epsilon(\tau)$ is not given exactly by a power-law for all $\tau$, it is difficult to quantify the ``error'' in the fitted parameters. One way to estimate the uncertainty, though, is to fit only a subset of the data points and see how the fitted parameters vary with the set. Doing so for the QA data points in the pure $p$-spin model, we find that $\epsilon_{\infty}$ decreases towards $\epsilon_{\textrm{th}}$ as we move the fitting window to larger $\tau$. This suggests that $\epsilon_{\infty}$ would likely agree better with $\epsilon_{\textrm{th}}$ were we to reach larger $\tau$ in our calculations.}\BibitemShut {Stop}%
\bibitem [{p_l()}]{p_limit_note}%
  \BibitemOpen
  \href@noop {} {}\bibinfo {note} {In order for a straightforward numerical solution of the integro-differential equations to be stable, one has to use a smaller timestep $\Delta t$ at larger $p$ (since the derivatives $f'[Q]$ and $f''[Q]$ become larger). This sets an upper limit to the values of $p$ that we can reasonably study.}\BibitemShut {Stop}%
\bibitem [{\citenamefont {Neal}(2011)}]{Neal2011MCMC}%
  \BibitemOpen
  \bibfield  {author} {\bibinfo {author} {\bibfnamefont {R.~M.}\ \bibnamefont {Neal}},\ }\bibfield  {title} {\bibinfo {title} {{MCMC} using {H}amiltonian dynamics},\ }in\ \href@noop {} {\emph {\bibinfo {booktitle} {Handbook of Markov Chain Monte Carlo}}}\ (\bibinfo  {publisher} {CRC Press},\ \bibinfo {year} {2011})\BibitemShut {NoStop}%
\bibitem [{\citenamefont {Betancourt}\ \emph {et~al.}(2017)\citenamefont {Betancourt}, \citenamefont {Byrne}, \citenamefont {Livingstone},\ and\ \citenamefont {Girolami}}]{Betancourt2017Geometric}%
  \BibitemOpen
  \bibfield  {author} {\bibinfo {author} {\bibfnamefont {M.}~\bibnamefont {Betancourt}}, \bibinfo {author} {\bibfnamefont {S.}~\bibnamefont {Byrne}}, \bibinfo {author} {\bibfnamefont {S.}~\bibnamefont {Livingstone}},\ and\ \bibinfo {author} {\bibfnamefont {M.}~\bibnamefont {Girolami}},\ }\bibfield  {title} {\bibinfo {title} {The geometric foundations of {H}amiltonian {M}onte {C}arlo},\ }\href@noop {} {\bibfield  {journal} {\bibinfo  {journal} {Bernoulli}\ }\textbf {\bibinfo {volume} {23}},\ \bibinfo {pages} {2257} (\bibinfo {year} {2017})}\BibitemShut {NoStop}%
\bibitem [{vec()}]{vector_spin_note}%
  \BibitemOpen
  \href@noop {} {}\bibinfo {note} {For the spherical $p$-spin models considered here, it is clear what the analogous classical algorithm to QA is: use Eq.~\eqref{eq:Hamiltonian_dynamics} but with classical $X_j$ and $P_j$ instead. Yet as discussed above, present-day quantum annealers are better suited to Ising models. For them, the obvious classical analogue is to replace the qubits by classical vector spins, but this immediately fails --- the local $z$-field on each spin vanishes in the initial state, and thus the spins remain frozen in the $x$ direction at all times. One would have to perturb the initial state to get any non-trivial dynamics, but how this perturbation would affect the asymptotic energy $\epsilon_{\infty}$ and power-law exponent $\alpha$ is unclear.}\BibitemShut {Stop}%
\end{thebibliography}%


\clearpage

\onecolumngrid
\makeatletter
\setcounter{MaxMatrixCols}{10}

\begin{center} 
{\large \bf Supplemental material for ``Reaching states below the threshold energy in spin glasses via quantum annealing''}
\end{center}

\setcounter{secnumdepth}{2} 
\renewcommand{\thesection}{S\arabic{section}} 
\renewcommand{\thesubsection}{\Alph{subsection}}

\setcounter{equation}{0}
\renewcommand{\theequation}{S\arabic{equation}}
\setcounter{table}{0}
\renewcommand{\thetable}{S\arabic{table}}
\setcounter{figure}{0}
\renewcommand{\thefigure}{S\arabic{figure}}

In this supplemental material, we derive and discuss the numerical solution of the integro-differential equations for the correlation and response functions of the spherical mixed $p$-spin model under quantum annealing (QA).
The Hamiltonian is
\begin{equation} \label{eq:quantum_annealing_p_spin_Hamiltonian}
H(t) = s_J(t) \sum_{p=2}^{\infty} \sqrt{a_p} \sum_{j_1 < \cdots < j_p} J_{j_1 \cdots j_p} \hat{X}_{j_1} \cdots \hat{X}_{j_p} + \frac{1}{2 s_K(t)} \sum_j \hat{P}_j^2 + \frac{z(t)}{2} \sum_j \hat{X}_j^2,
\end{equation}
where each $(\hat{X}_j, \hat{P}_j)$ is a canonically conjugate position-momentum pair.
QA corresponds to setting $s_J(t) = t/\tau$ and $s_K(t) = (1 - t/\tau)^{-1}$, but we will find it useful to allow for arbitrary time-dependent coefficients.
The final term exists to enforce the spherical constraint $\sum_j \langle \hat{X}_j(t)^2 \rangle = N$ at all times.
As initial state, we use the product state in which each coordinate has the Gaussian wavefunction $\psi_0(x_j) \propto \exp{[-x_j^2/4]}$ --- note that this wavefunction is indeed the ground state of the initial Hamiltonian, which is (since $s_J(0) = 0$) simply a harmonic oscillator whose stiffness is chosen so that the ground state has $\langle \hat{X}_j^2 \rangle = 1$.
Lastly, each $J_{j_1 \cdots j_p}$ is an independent Gaussian random variable of mean zero and variance $p!/2N^{p-1}$ (whereas the coefficients $a_p$ are fixed parameters).

Following the standard derivation of a path integral from a Hamiltonian, Eq.~\eqref{eq:quantum_annealing_p_spin_Hamiltonian} yields the Keldysh path integral
\begin{equation} \label{eq:Keldysh_path_integral}
\begin{aligned}
\mathcal{Z} \equiv \int \mathcal{D}x &\exp{\left[ i \int \textrm{d}t \left( \frac{s_K^+(t)}{2} \sum_j \big( \partial_t x_j^+(t) \big)^2 - \frac{z(t)}{2} \sum_j x_j^+(t)^2 - s_J^+(t) \sum_p \sqrt{a_p} \sum_{j_1 < \cdots < j_p} J_{j_1 \cdots j_p} x_{j_1}^+(t) \cdots x_{j_p}^+(t) \right) \right]} \\
\cdot &\exp{\left[ -i \int \textrm{d}t \left( \frac{s_K^-(t)}{2} \sum_j \big( \partial_t x_j^-(t) \big)^2 - \frac{z(t)}{2} \sum_j x_j^-(t)^2 - s_J^-(t) \sum_p \sqrt{a_p} \sum_{j_1 < \cdots < j_p} J_{j_1 \cdots j_p} x_{j_1}^-(t) \cdots x_{j_p}^-(t) \right) \right]} \\
\cdot &\exp{\left[ -\frac{1}{4} \sum_j \Big( x_j^+(0)^2 + x_j^-(0)^2 \Big) \right]}.
\end{aligned}
\end{equation}
Some comments are in order:
\begin{itemize}
\item As usual, we have separate integration variables for the forward ($+$) and backward ($-$) branches.
\item The final line is the initial state $\prod_j \langle x_j^+(0) | \psi_0 \rangle \langle \psi_0 | x_j^-(0) \rangle$.
Note that we do \textit{not} fix $x_j^+(0) = x_j^-(0)$.
\item We use independent coefficients $s_J^{\pm}(t)$ and $s_K^{\pm}(t)$ for the forward and backward branches, at least for now, so that we can take appropriate functional derivatives to determine the average potential and kinetic energies.
In particular, the average potential energy density $\epsilon(t)$ is given simply by
\begin{equation} \label{eq:potential_energy_initial}
\epsilon(t) = \frac{i}{N} \frac{\delta \mathbb{E}[\mathcal{Z}]}{\delta s_J^+(t)}.
\end{equation}
\item Although we use the standard continuum notation for convenience, we are really integrating over the discrete set $\{ x_j^{\pm}(0), x_j^{\pm}(\Delta t), x_j^{\pm}(2 \Delta t), \cdots \}$, where the path integral becomes exact in the limit $\Delta t \rightarrow 0$ but we will have to use finite $\Delta t$ in order to solve the resulting equations numerically.
The integral $\int \textrm{d}t$ is short-hand for the discrete sum $\sum_t \Delta t$, and the derivative $\partial_t x_j^{\pm}(t)$ is short-hand for the finite difference $[x_j^{\pm}(t + \Delta t) - x_j^{\pm}(t)]/\Delta t$.
\end{itemize}

The key order parameters are the correlation function $C(t, t')$ and response function $R(t, t')$.
These are given by
\begin{equation} \label{eq:correlation_response_definitions}
C(t, t') \equiv \frac{1}{2N} \sum_j \big< \hat{X}_j(t) \hat{X}_j(t') + \hat{X}_j(t') \hat{X}_j(t) \big>, \qquad R(t, t') \equiv \frac{i}{N} \Theta(t - t') \sum_j \big< \big[ \hat{X}_j(t), \hat{X}_j(t') \big] \big>.
\end{equation}
We can express these in terms of expectation values of the classical variables $x_j^{\pm}(t)$ (under the path integral) by choosing the branches appropriately.
For our purposes, the inverse relationships will be more useful, which one can easily confirm using the definitions in Eq.~\eqref{eq:correlation_response_definitions} and keeping track of time-ordering:
\begin{equation} \label{eq:classical_operator_expectation_relationships}
\begin{aligned}
\frac{1}{N} \sum_j \big< x_j^+(t) x_j^+(t') \big> &= C(t, t') - \frac{i}{2} \Big( R(t, t') + R(t', t) \Big), \\
\frac{1}{N} \sum_j \big< x_j^+(t) x_j^-(t') \big> &= C(t, t') + \frac{i}{2} \Big( R(t, t') - R(t', t) \Big), \\
\frac{1}{N} \sum_j \big< x_j^-(t) x_j^+(t') \big> &= C(t, t') - \frac{i}{2} \Big( R(t, t') - R(t', t) \Big), \\
\frac{1}{N} \sum_j \big< x_j^-(t) x_j^-(t') \big> &= C(t, t') + \frac{i}{2} \Big( R(t, t') + R(t', t) \Big).
\end{aligned}
\end{equation}

The first step is to average over the coefficients $J_{j_1 \cdots j_p}$.
Using $\sigma \in \{ +1, -1 \}$ to denote the branch, note that
\begin{equation} \label{eq:exponential_average}
\begin{aligned}
&\mathbb{E} \exp{\left[ -i \int \textrm{d}t \sum_{\sigma} \sigma s_J^{\sigma}(t) \sum_p \sqrt{a_p} \sum_{j_1 < \cdots < j_p} J_{j_1 \cdots j_p} x_{j_1}^{\sigma}(t) \cdots x_{j_p}^{\sigma}(t) \right]} \\
&\qquad \qquad \qquad = \exp{\left[ -\sum_p \frac{p! a_p}{4N^{p-1}} \int \textrm{d}t \textrm{d}t' \sum_{\sigma \sigma'} \sigma \sigma' s_J^{\sigma}(t) s_J^{\sigma'}(t') \sum_{j_1 < \cdots < j_p} x_{j_1}^{\sigma}(t) x_{j_1}^{\sigma'}(t') \cdots x_{j_p}^{\sigma}(t) x_{j_p}^{\sigma'}(t') \right]} \\
&\qquad \qquad \qquad = \exp{\left[ -\sum_p \frac{a_p}{4N^{p-1}} \int \textrm{d}t \textrm{d}t' \sum_{\sigma \sigma'} \sigma \sigma' s_J^{\sigma}(t) s_J^{\sigma'}(t') \sum_{j_1 \neq \cdots \neq j_p} x_{j_1}^{\sigma}(t) x_{j_1}^{\sigma'}(t') \cdots x_{j_p}^{\sigma}(t) x_{j_p}^{\sigma'}(t') \right]} \\
&\qquad \qquad \qquad \sim \exp{\left[ -\sum_p \frac{N a_p}{4} \int \textrm{d}t \textrm{d}t' \sum_{\sigma \sigma'} \sigma \sigma' s_J^{\sigma}(t) s_J^{\sigma'}(t') \bigg( \frac{1}{N} \sum_j x_j^{\sigma}(t) x_j^{\sigma'}(t') \bigg)^p \right]} \\
&\qquad \qquad \qquad = \exp{\left[ -\frac{N}{4} \int \textrm{d}t \textrm{d}t' \sum_{\sigma \sigma'} \sigma \sigma' s_J^{\sigma}(t) s_J^{\sigma'}(t') f \bigg[ \frac{1}{N} \sum_j x_j^{\sigma}(t) x_j^{\sigma'}(t') \bigg] \right]},
\end{aligned}
\end{equation}
recalling the definition $f[Q] \equiv \sum_p a_p Q^p$.
Thus the disorder-averaged path integral is
\begin{equation} \label{eq:average_Keldysh_path_integral}
\begin{aligned}
\mathbb{E}[\mathcal{Z}] = \int \mathcal{D}x \, &\exp{\left[ i \int \textrm{d}t \sum_{\sigma} \sigma \left( \frac{s_K^{\sigma}(t)}{2} \sum_j \big( \partial_t x_j^{\sigma}(t) \big)^2 - \frac{z(t)}{2} \sum_j x_j^{\sigma}(t)^2 \right) - \frac{1}{4} \sum_{j \sigma} x_j^{\sigma}(0)^2 \right]} \\
\cdot &\exp{\left[ -\frac{N}{4} \int \textrm{d}t \textrm{d}t' \sum_{\sigma \sigma'} \sigma \sigma' s_J^{\sigma}(t) s_J^{\sigma'}(t') f \bigg[ \frac{1}{N} \sum_j x_j^{\sigma}(t) x_j^{\sigma'}(t') \bigg] \right]}.
\end{aligned}
\end{equation}
The only piece that doesn't factor across $j$ is the lower line, involving $N^{-1} \sum_j x_j^{\sigma}(t) x_j^{\sigma'}(t')$.
Thus introduce additional integration variables $Q^{\sigma \sigma'}(t, t')$ and $\lambda^{\sigma \sigma'}(t, t')$ (for all $t > t'$ and all $\sigma, \sigma'$) via the identity
\begin{equation} \label{eq:fat_unity_identity}
\begin{aligned}
1 &= \int \mathcal{D}Q \prod_{t > t'} \prod_{\sigma \sigma'} \delta \bigg( Q^{\sigma \sigma'}(t, t') - \frac{1}{N} \sum_j x_j^{\sigma}(t) x_j^{\sigma'}(t') \bigg) \\
&= \int \mathcal{D}Q \mathcal{D}\lambda \exp{\left[ i N \int_{t > t'} \textrm{d}t \textrm{d}t' \sum_{\sigma \sigma'} \sigma \sigma' \lambda^{\sigma \sigma'}(t, t') \bigg( Q^{\sigma \sigma'}(t, t') - \frac{1}{N} \sum_j x_j^{\sigma}(t) x_j^{\sigma'}(t') \bigg) \right]}.
\end{aligned}
\end{equation}
Then we have
\begin{equation} \label{eq:average_Keldysh_path_integral_factored}
\begin{aligned}
\mathbb{E}[\mathcal{Z}] &= \int \mathcal{D}Q \mathcal{D}\lambda \exp{\left[ -\frac{N}{2} \int_{t > t'} \textrm{d}t \textrm{d}t' \sum_{\sigma \sigma'} \sigma \sigma' s_J^{\sigma}(t) s_J^{\sigma'}(t') f \big[ Q^{\sigma \sigma'}(t, t') \big] + iN \int_{t > t'} \textrm{d}t \textrm{d}t' \sum_{\sigma \sigma'} \sigma \sigma' \lambda^{\sigma \sigma'}(t, t') Q^{\sigma \sigma'}(t, t') \right]} \\
&\qquad \qquad \cdot \prod_j \int \mathcal{D}x_j \exp{\left[ i \int \textrm{d}t \sum_{\sigma} \sigma \left( \frac{s_K^{\sigma}(t)}{2} \big( \partial_t x_j^{\sigma}(t) \big)^2 - \frac{z(t)}{2} x_j^{\sigma}(t)^2 \right) - \frac{1}{4} \sum_{\sigma} x_j^{\sigma}(0)^2 \right]} \\
&\qquad \qquad \qquad \qquad \cdot \exp{\left[ -i \int_{t > t'} \textrm{d}t \textrm{d}t' \sum_{\sigma \sigma'} \sigma \sigma' \lambda^{\sigma \sigma'}(t, t') x_j^{\sigma}(t) x_j^{\sigma'}(t') \right]}.
\end{aligned}
\end{equation}
Evaluating the Gaussian integrals over $x$ gives that $\mathbb{E}[\mathcal{Z}] = \int \mathcal{D}Q \mathcal{D}\lambda \exp{[NS]}$, where the action $S[Q, \lambda]$ is
\begin{equation} \label{eq:Keldysh_action}
\begin{aligned}
S[Q, \lambda] = &-\frac{1}{2} \int_{t > t'} \textrm{d}t \textrm{d}t' \sum_{\sigma \sigma'} \sigma \sigma' \Big( s_J^{\sigma}(t) s_J^{\sigma'}(t') f \big[ Q^{\sigma \sigma'}(t, t') \big] - 2i \lambda^{\sigma \sigma'}(t, t') Q^{\sigma \sigma'}(t, t') \Big) \\
&- \frac{1}{2} \log{\textrm{Det}} \bigg[ \sigma \delta(t - t') \delta_{\sigma \sigma'} \Big( \partial_t s_K^{\sigma} \partial_t + z(t) \Big) - \frac{i}{2} \delta(t) \delta(t') \delta_{\sigma \sigma'} + \sigma \sigma' \lambda^{\sigma \sigma'}(t, t') \bigg].
\end{aligned}
\end{equation}
Note that we have neglected terms which are independent of $Q$ and $\lambda$ (this is why we have been glib about the normalization of the path integral).
For reference, the proper discretization of the matrix $\partial_t s_K^{\sigma} \partial_t$ in the lower line is
\begin{equation} \label{eq:second_derivative_explicit}
\partial_t s_K^{\sigma} \partial_t = \frac{1}{\Delta t^2} \begin{pmatrix} \vphantom{\ddots}-s_K^{\sigma}(0) & s_K^{\sigma}(0) & & & \\ \vphantom{\ddots}s_K^{\sigma}(0) & -s_K^{\sigma}(0) - s_K^{\sigma}(\Delta t) & s_K^{\sigma}(\Delta t) & & \\ \vphantom{\ddots} & s_K^{\sigma}(\Delta t) & -s_K^{\sigma}(\Delta t) - s_K^{\sigma}(2 \Delta t) & s_K^{\sigma}(2 \Delta t) & \\ & & s_K^{\sigma}(2 \Delta t) & \ddots & \ddots \\ & & & \ddots & \end{pmatrix}.
\end{equation}
The remaining path integral over $Q$ and $\lambda$ can be evaluated by saddle-point approximation at large $N$.
The saddle-point equations are
\begin{equation} \label{eq:saddle_point_equations}
\begin{gathered}
2i \lambda^{\sigma \sigma'}(t, t') = s_J^{\sigma}(t) s_J^{\sigma'}(t') f' \big[ Q^{\sigma \sigma'}(t, t') \big], \\
iQ^{\sigma \sigma'}(t, t') = \bigg[ \sigma \delta(t - t') \delta_{\sigma \sigma'} \Big( \partial_t s_K^{\sigma} \partial_t + z(t) \Big) - \frac{i}{2} \delta(t) \delta(t') \delta_{\sigma \sigma'} + \sigma \sigma' \lambda^{\sigma \sigma'}(t, t') \bigg]^{-1},
\end{gathered}
\end{equation}
where the lower line refers to the inverse of the matrix in brackets (as opposed to the reciprocal of a matrix element).
These can be expressed more compactly as the set of integro-differential equations
\begin{equation} \label{eq:integro_differential_equations}
\Big( \partial_t s_K^{\sigma} \partial_t + z(t) - \frac{i \sigma}{2} \delta(t) \Big) Q^{\sigma \sigma'}(t, t') - \frac{i s_J^{\sigma}(t)}{2} \int \textrm{d}t'' s_J^{\sigma''}(t'') \sum_{\sigma''} \sigma'' f' \big[ Q^{\sigma \sigma''}(t, t'') \big] Q^{\sigma'' \sigma'}(t'', t') = -i \sigma \delta(t - t') \delta_{\sigma \sigma'}.
\end{equation}

At this point, we can differentiate the path integral with respect to $s_J^+(t)$ to determine the average potential energy (see Eq.~\eqref{eq:potential_energy_initial}).
This brings down a factor of $\delta S/\delta s_J^+(t)$ inside the path integral, which evaluates to the saddle-point value of $\delta S/\delta s_J^+(t)$ according to the saddle-point approximation.
Thus
\begin{equation} \label{eq:potential_energy_explicit}
\epsilon(t) = -\frac{i}{2} \int \textrm{d}t'' \sum_{\sigma''} \sigma'' s_J^{\sigma''}(t'') f \big[ Q^{+ \sigma''}(t, t'') \big],
\end{equation}
where $Q^{\sigma \sigma'}(t, t')$ is determined by solving Eq.~\eqref{eq:integro_differential_equations}.
Now that we have this expression, we can safely set $s_J^+(t) = s_J^-(t)$ and $s_K^+(t) = s_K^-(t)$ --- we will henceforth neglect the superscripts on $s_J(t)$ and $s_K(t)$ (although still keep them independent of each other for generality).

Since we first introduced $Q^{\sigma \sigma'}(t, t')$ via a $\delta$-function fixing it to $N^{-1} \sum_j x_j^{\sigma}(t) x_j^{\sigma'}(t')$, the saddle-point value of $Q^{\sigma \sigma'}(t, t')$ equals the expectation value of $N^{-1} \sum_j x_j^{\sigma}(t) x_j^{\sigma'}(t')$.
Thus $Q^{\sigma \sigma'}(t, t')$ should have the same expression in terms of the correlation and response functions as in Eq.~\eqref{eq:classical_operator_expectation_relationships}, which we can summarize by writing
\begin{equation} \label{eq:order_parameter_expression}
Q^{\sigma \sigma'}(t, t') = C(t, t') - \frac{i \sigma'}{2} R(t, t') - \frac{i \sigma}{2} R(t', t).
\end{equation}
Keep in mind that due to causality, $R(t, t') = 0$ for $t \leq t'$, and thus one of the latter two terms will automatically vanish for any choice of times (while both vanish for $t = t'$).
Furthermore, although we only defined $Q^{\sigma \sigma'}(t, t')$ for $t > t'$, the ansatz in Eq.~\eqref{eq:order_parameter_expression} can be extended to $t < t'$ by symmetry ($Q^{\sigma' \sigma}(t', t) = Q^{\sigma \sigma'}(t, t')$), and the spherical constraint means that we should set $Q^{\sigma \sigma'}(t, t) = 1$ (equivalently $C(t, t) = 1$).

One reassuring consequence of this ansatz is that Eq.~\eqref{eq:integro_differential_equations} respects the causality structure, i.e., the integral over $t''$ has no contribution from $t'' \geq \max\{t, t'\}$ and so $Q^{\sigma \sigma'}(t, t')$ is fully determined by its values at previous times.
To see this, simply note that $Q^{\sigma \sigma'}(t, t')$ is independent of $\sigma$ for $t \geq t'$ and independent of $\sigma'$ for $t' \geq t$.
Thus if $t'' \geq \max\{t, t'\}$ in Eq.~\eqref{eq:integro_differential_equations}, then the summand is proportional to $\sigma''$ and so the sum over $\sigma''$ vanishes.

To rewrite Eq.~\eqref{eq:integro_differential_equations} in terms of equations for $C(t, t')$ and $R(t, t')$, separate the real and imaginary parts.
We have to consider various cases individually.
First suppose $t > t'$.
The real part of Eq.~\eqref{eq:integro_differential_equations} is
\begin{equation} \label{eq:correlation_equation_initial}
\begin{aligned}
&\Big( \partial_t s_K \partial_t + z(t) \Big) C(t, t') \\
&\qquad + \frac{s_J(t)}{2} \int \textrm{d}t'' s_J(t'') \sum_{\sigma''} \sigma'' \textrm{Im} f' \Big[ C(t, t'') - \frac{i \sigma''}{2} R(t, t'') \Big] \Big( C(t'', t') - \frac{i \sigma'}{2} R(t'', t') - \frac{i \sigma''}{2} R(t', t'') \Big) = 0,
\end{aligned}
\end{equation}
and the imaginary part is
\begin{equation} \label{eq:response_equation_initial}
\begin{aligned}
&-\frac{\sigma'}{2} \Big( \partial_t s_K \partial_t + z(t) \Big) R(t, t') \\
&\qquad - \frac{s_J(t)}{2} \int \textrm{d}t'' s_J(t'') \sum_{\sigma''} \sigma'' \textrm{Re} f' \Big[ C(t, t'') - \frac{i \sigma''}{2} R(t, t'') \Big] \Big( C(t'', t') - \frac{i \sigma'}{2} R(t'', t') - \frac{i \sigma''}{2} R(t', t'') \Big) = 0.
\end{aligned}
\end{equation}
Note that $f'[Q] \equiv \sum_p p a_p Q^{p-1}$ is a polynomial in $Q$, and its argument involves $i$ and $\sigma''$ only through the product $i \sigma''$.
Thus its real part is independent of $\sigma''$ and its imaginary part is proportional to $\sigma''$, i.e.,
\begin{equation} \label{eq:f_polynomial_form}
f' \Big[ C(t, t'') - \frac{i \sigma''}{2} R(t, t'') \Big] = \textrm{Re} f' \Big[ C(t, t'') - \frac{i}{2} R(t, t'') \Big] + i \sigma'' \textrm{Im} f' \Big[ C(t, t'') - \frac{i}{2} R(t, t'') \Big].
\end{equation}
Inserting into Eqs.~\eqref{eq:correlation_equation_initial} and~\eqref{eq:response_equation_initial}, many terms do not survive taking the real/imaginary part and summing over $\sigma''$.
We are left with
\begin{equation} \label{eq:correlation_equation_simplified}
\Big( \partial_t s_K \partial_t + z(t) \Big) C(t, t') + s_J(t) \int \textrm{d}t'' s_J(t'') \textrm{Im} f' \Big[ C(t, t'') - \frac{i}{2} R(t, t'') \Big] \Big( C(t', t'') - \frac{i}{2} R(t', t'') \Big) = 0,
\end{equation}
and (after canceling factors of $-\sigma'/2$)
\begin{equation} \label{eq:response_equation_simplified}
\Big( \partial_t s_K \partial_t + z(t) \Big) R(t, t') + s_J(t) \int \textrm{d}t'' s_J(t'') \textrm{Im} f' \Big[ C(t, t'') - \frac{i}{2} R(t, t'') \Big] R(t'', t') = 0.
\end{equation}
Note that the $\sigma$ and $\sigma'$ indices have indeed dropped out, leaving us with just two equations for two unknowns.

Next suppose $t = t' > 0$.
Eq.~\eqref{eq:correlation_equation_simplified} still holds for the real part, and since $C(t, t) = 1$ and $C(t - \Delta t, t) = C(t, t - \Delta t)$, this gives $C(t + \Delta t, t)$ in terms of $C(t, t - \Delta t)$ once we use the explicit form of $\partial_t s_K \partial_t$ (Eq.~\eqref{eq:second_derivative_explicit}):
\begin{equation} \label{eq:curvature_equation_v1}
\begin{aligned}
&\frac{1}{\Delta t^2} \Big( s_K(t) C(t + \Delta t, t) - s_K(t) - s_K(t - \Delta t) + s_K(t - \Delta t) C(t, t - \Delta t) \Big) + z(t) \\
&\qquad \qquad \qquad \qquad + s_J(t) \int \textrm{d}t'' s_J(t'') \textrm{Im} f' \Big[ C(t, t'') - \frac{i}{2} R(t, t'') \Big] \Big( C(t, t'') - \frac{i}{2} R(t, t'') \Big) = 0.
\end{aligned}
\end{equation}
It is convenient to define $A(t)$ by writing $C(t + \Delta t, t) \equiv 1 - A(t) \Delta t^2$.
Then Eq.~\eqref{eq:curvature_equation_v1} gives $A(t)$ in terms of $A(t - \Delta t)$:
\begin{equation} \label{eq:curvature_equation_v2}
s_K(t) A(t) = -s_K(t - \Delta t) A(t - \Delta t) + z(t) + s_J(t) \int \textrm{d}t'' s_J(t'') \textrm{Im} f' \Big[ C(t, t'') - \frac{i}{2} R(t, t'') \Big] \Big( C(t, t'') - \frac{i}{2} R(t, t'') \Big).
\end{equation}
For the imaginary part of Eq.~\eqref{eq:integro_differential_equations}, the integral in fact vanishes, but the right-hand side is now non-zero.
Using the explicit form of $\partial_t s_K \partial_t$ and Eq.~\eqref{eq:order_parameter_expression}, we have that
\begin{equation} \label{eq:initial_response_equation}
-\frac{\sigma' s_K(t)}{2 \Delta t^2} R(t + \Delta t, t) - \frac{\sigma s_K(t - \Delta t)}{2 \Delta t^2} R(t, t - \Delta t) = -\frac{\sigma}{\Delta t} \delta_{\sigma \sigma'}.
\end{equation}
One can confirm that by setting $R(t + \Delta t, t) = \Delta t/s_K(t)$ and $R(t, t - \Delta t) = \Delta t/s_K(t - \Delta t)$, this equation is indeed satisfied for all $\sigma$ and $\sigma'$.
Thus together with $C(t, t) = 1$ and $R(t, t) = 0$, Eqs.~\eqref{eq:curvature_equation_v2} and~\eqref{eq:initial_response_equation} give us the initial conditions for $C(t, t')$ and $R(t, t')$.

Lastly suppose $t = t' = 0$, since Eq.~\eqref{eq:integro_differential_equations} has an additional term and $\partial_t s_K \partial_t$ takes a different form.
The equation becomes
\begin{equation} \label{eq:initial_integro_differential_equation}
\frac{s_K(0)}{\Delta t^2} \Big( Q^{\sigma \sigma'}(\Delta t, 0) - 1 \Big) + z(t) - \frac{i \sigma}{2 \Delta t} = -\frac{i \sigma}{\Delta t} \delta_{\sigma \sigma'}.
\end{equation}
The real part amounts to $s_K(0) A(0) = z(0)$, and the imaginary part is again solved by setting $R(\Delta t, 0) = \Delta t/s_K(0)$, consistent with what we found above.

It remains only to determine $z(t)$, which we do by requiring that the initial condition $C(t, t) = 1$ be compatible with Eq.~\eqref{eq:integro_differential_equations}.
Again take the real part, but now at $t' = t + \Delta t$:
\begin{equation} \label{eq:Lagrange_multiplier_equation_1}
\begin{aligned}
&\frac{1}{\Delta t^2} \Big( s_K(t) - \big[ s_K(t) + s_K(t - \Delta t) \big] C(t + \Delta t, t) + s_K(t - \Delta t) C(t + \Delta t, t - \Delta t) \Big) + z(t) C(t + \Delta t, t) \\
&\qquad \qquad \qquad \qquad \qquad + s_J(t) \int \textrm{d}t'' s_J(t'') \textrm{Im} f' \Big[ C(t, t'') - \frac{i}{2} R(t, t'') \Big] \Big( C(t + \Delta t, t'') - \frac{i}{2} R(t + \Delta t, t'') \Big) = 0,
\end{aligned}
\end{equation}
where we used that $C(t, t') = C(t', t)$.
Meanwhile, recall Eq.~\eqref{eq:curvature_equation_v1}:
\begin{equation} \label{eq:Lagrange_multiplier_equation_2}
\begin{aligned}
&\frac{1}{\Delta t^2} \Big( s_K(t) C(t + \Delta t, t) - s_K(t) - s_K(t - \Delta t) + s_K(t - \Delta t) C(t, t - \Delta t) \Big) + z(t) \\
&\qquad \qquad \qquad \qquad + s_J(t) \int \textrm{d}t'' s_J(t'') \textrm{Im} f' \Big[ C(t, t'') - \frac{i}{2} R(t, t'') \Big] \Big( C(t, t'') - \frac{i}{2} R(t, t'') \Big) = 0,
\end{aligned}
\end{equation}
and set $t' = t - \Delta t$ in Eq.~\eqref{eq:correlation_equation_simplified}:
\begin{equation} \label{eq:Lagrange_multiplier_equation_3}
\begin{aligned}
&\frac{1}{\Delta t^2} \Big( s_K(t) C(t + \Delta t, t - \Delta t) - \big[ s_K(t) + s_K(t - \Delta t) \big] C(t, t - \Delta t) + s_K(t - \Delta t) \Big) + z(t) C(t, t - \Delta t) \\
&\qquad \qquad \qquad \qquad + s_J(t) \int \textrm{d}t'' s_J(t'') \textrm{Im} f' \Big[ C(t, t'') - \frac{i}{2} R(t, t'') \Big] \Big( C(t - \Delta t, t'') - \frac{i}{2} R(t - \Delta t, t'') \Big) = 0.
\end{aligned}
\end{equation}
These equations determine $z(t)$ in terms of the correlation function at previous times.
Take $s_K(t)$ times Eq.~\eqref{eq:Lagrange_multiplier_equation_1}, $[s_K(t) + s_K(t - \Delta t)]$ times Eq.~\eqref{eq:Lagrange_multiplier_equation_2}, $-s_K(t - \Delta t)$ times Eq.~\eqref{eq:Lagrange_multiplier_equation_3}, and add the three together:
\begin{equation} \label{eq:Lagrange_multiplier_equation_combined}
\begin{aligned}
&\Big[ s_K(t) C(t + \Delta t, t) + s_K(t) + s_K(t - \Delta t) - s_K(t - \Delta t) C(t, t - \Delta t) \Big] z(t) \\
&\qquad \qquad = 2 s_K(t - \Delta t) \big[ s_K(t) + s_K(t - \Delta t) \big] \frac{1 - C(t, t - \Delta t)}{\Delta t^2} \\
& \qquad \qquad \qquad - s_J(t) \int \textrm{d}t'' s_J(t'') \textrm{Im} f' \Big[ C(t, t'') - \frac{i}{2} R(t, t'') \Big] \bigg[ s_K(t) \Big( C(t + \Delta t, t'') - \frac{i}{2} R(t + \Delta t, t'') \Big) \\
&\qquad \qquad \qquad \qquad \qquad \qquad \qquad \qquad \qquad \qquad \qquad \qquad \qquad + \big[ s_K(t) + s_K(t - \Delta t) \big] \Big( C(t, t'') - \frac{i}{2} R(t, t'') \Big) \\
&\qquad \qquad \qquad \qquad \qquad \qquad \qquad \qquad \qquad \qquad \qquad \qquad \qquad - s_K(t - \Delta t) \Big( C(t - \Delta t, t'') - \frac{i}{2} R(t - \Delta t, t'') \Big) \bigg].
\end{aligned}
\end{equation}
We only need $z(t)$ to zeroth order in $\Delta t$ to integrate Eqs.~\eqref{eq:correlation_equation_simplified} and~\eqref{eq:response_equation_simplified}.
Eq.~\eqref{eq:Lagrange_multiplier_equation_combined} simplifies significantly to zeroth order, again writing $C(t, t - \Delta t) = 1 - A(t - \Delta t) \Delta t^2$:
\begin{equation} \label{eq:Lagrange_multiplier_equation_zeroth_order}
z(t) = 2 s_K(t) A(t - \Delta t) - s_J(t) \int \textrm{d}t'' s_J(t'') \textrm{Im} f' \Big[ C(t, t'') - \frac{i}{2} R(t, t'') \Big] \Big( C(t, t'') - \frac{i}{2} R(t, t'') \Big).
\end{equation}
However, we need $z(t)$ to first order in $\Delta t$ to integrate Eq.~\eqref{eq:curvature_equation_v2}.
Inserting Eq.~\eqref{eq:Lagrange_multiplier_equation_combined} into Eq.~\eqref{eq:curvature_equation_v2} and dropping terms of $O(\Delta t^2)$ gives, after some algebra,
\begin{equation} \label{eq:curvature_equation_simplified}
A(t) = \frac{s_K(t - \Delta t)^2}{s_K(t)^2} A(t - \Delta t) - \frac{s_J(t) \Delta t}{s_K(t)} \int \textrm{d}t'' s_J(t'') \partial_t \textrm{Im} f \Big[ C(t, t'') - \frac{i}{2} R(t, t'') \Big].
\end{equation}

Once again, we have to treat $z(0)$ separately, which we do by setting $t = 0$ and $t' = \Delta t$ in Eq.~\eqref{eq:integro_differential_equations}.
The real part is
\begin{equation} \label{eq:initial_Lagrange_multiplier_equation}
\frac{s_K(0)}{\Delta t^2} \Big( 1 - C(\Delta t, 0) \Big) + z(0) C(\Delta t, 0) - \frac{1}{4 \Delta t} R(\Delta t, 0) - \frac{s_J(0)^2 f'[1] \Delta t}{2} R(\Delta t, 0) = 0.
\end{equation}
Since $R(\Delta t, 0) = \Delta t/s_K(0)$ and $z(0) = s_K(0) A(0)$ (see below Eq.~\eqref{eq:initial_integro_differential_equation}), dropping terms of $O(\Delta t^2)$ gives us the initial value $A(0) = 1/8 s_K(0)^2$ (and thus $z(0) = 1/8 s_K(0)$).

Lastly, do not forget about the potential energy, which is what we're ultimately after.
Returning to Eq.~\eqref{eq:potential_energy_explicit} and using Eq.~\eqref{eq:order_parameter_expression}, we have that
\begin{equation} \label{eq:potential_energy_result}
\epsilon(t) = \int \textrm{d}t'' s_J(t'') \textrm{Im} f \Big[ C(t, t'') - \frac{i}{2} R(t, t'') \Big].
\end{equation}

This gives us everything that we need.
While we have made some small-$\Delta t$ approximations, we can solve the equations at finite $\Delta t$ and obtain results that become exact as $\Delta t \rightarrow 0$.
In principle the approximations could have led to numerically unstable equations, but these appear to be stable as long as $\Delta t$ is sufficiently small.
To summarize the numerical procedure:
\begin{tcolorbox}[colback=boxback]
Denote $Q(t, t') \equiv C(t, t') - i R(t, t')/2$, and define $f[Q] \equiv \sum_p a_p Q^p$.
Use initial conditions $C(t, t) = 1$ and $R(t, t) = 0$, together with (writing $C(t + \Delta t, t) = 1 - A(t) \Delta t^2$)
\begin{equation} \label{eq:curvature_summary}
A(t) = \frac{s_K(t - \Delta t)^2}{s_K(t)^2} A(t - \Delta t) - \frac{s_J(t) \Delta t}{s_K(t)} \int_0^t \textrm{d}t'' s_J(t'') \partial_t \textrm{Im} f \big[ Q(t, t'') \big],
\end{equation}
and $R(t + \Delta t, t) = \Delta t/s_K(t)$.
We begin with $z(0) = 1/8 s_K(0)$ and $A(0) = 1/8 s_K(0)^2$.
Once we've determined $C(t, t')$ and $R(t, t')$ for all $t' \leq t$, set
\begin{equation} \label{eq:Lagrange_multiplier_summary}
z(t) = 2 s_K(t) A(t - \Delta t) - s_J(t) \int_0^t \textrm{d}t'' s_J(t'') \textrm{Im} f' \big[ Q(t, t'') \big] Q(t, t''),
\end{equation}
then solve for $C(t + \Delta t, t')$ and $R(t + \Delta t, t')$:
\begin{equation} \label{eq:correlation_summary}
\begin{aligned}
\frac{1}{\Delta t^2} \Big( s_K(t) C(t + \Delta t, t') &- \big[ s_K(t) + s_K(t - \Delta t) \big] C(t, t') + s_K(t - \Delta t) C(t - \Delta t, t') \Big) + z(t) C(t, t') \\
&+ s_J(t) \int_0^t \textrm{d}t'' s_J(t'') \textrm{Im} f' \big[ Q(t, t'') \big] Q(t', t'') = 0,
\end{aligned}
\end{equation}
and
\begin{equation} \label{eq:response_summary}
\begin{aligned}
\frac{1}{\Delta t^2} \Big( s_K(t) R(t + \Delta t, t') &- \big[ s_K(t) + s_K(t - \Delta t) \big] R(t, t') + s_K(t - \Delta t) R(t - \Delta t, t') \Big) + z(t) R(t, t') \\
&+ s_J(t) \int_{t'}^t \textrm{d}t'' s_J(t'') \textrm{Im} f' \big[ Q(t, t'') \big] R(t'', t') = 0.
\end{aligned}
\end{equation}
The potential energy density is lastly given by
\begin{equation} \label{eq:potential_energy_summary}
\epsilon(t) = \int_0^t \textrm{d}t'' s_J(t'') \textrm{Im} f \big[ Q(t, t'') \big].
\end{equation}
For QA in particular, use $s_J(t) = t/\tau$ and $s_K(t) = (1 - t/\tau)^{-1}$.
\end{tcolorbox}

\end{document}